\documentclass[iop,twocolappendix]{emulateapj}
\usepackage{amsmath,mathtools}
\newcommand{\tf}{\theta_\mathrm{funnel}}
\newcommand{\Tf}{T_\mathrm{e,funnel}}

\begin{document}
\title{Variability in GRMHD simulations of Sgr~A$^*$: Implications for EHT closure phase observations}
\author{Lia Medeiros\altaffilmark{1,2}, Chi-kwan Chan\altaffilmark{1}, Feryal \"Ozel\altaffilmark{1}, Dimitrios Psaltis\altaffilmark{1}, \\Junhan Kim\altaffilmark{1}, Daniel Marrone\altaffilmark{1}, and Aleksander S{\c a}dowski\altaffilmark{3}}

\altaffiltext{1}{Steward Observatory and Department of Astronomy, University of Arizona, 933 N. Cherry Ave., Tucson, AZ 85721}
\altaffiltext{2}{Department of Physics, Broida Hall, University of California Santa Barbara, Santa Barbara, CA 93106}
\altaffiltext{3}{MIT Kavli Institute for Astrophysics and Space Research, 77 Massachusetts Ave, Cambridge, MA 02139}

\begin{abstract}
The observable quantities that carry the most information regarding
the structures of the images of black holes in the interferometric
observations with the Event Horizon Telescope are the closure phases
along different baseline triangles. We use long time span, high
cadence, GRMHD+radiative transfer models of Sgr~A$^*$ to investigate the
expected variability of closure phases in such observations. We find
that, in general, closure phases along small baseline triangles show
little variability, except in the cases when one of the triangle
vertices crosses one of a small regions of low visibility
amplitude. The closure phase variability increases with the size of
the baseline triangle, as larger baselines probe the small-scale
structures of the images, which are highly variable. On average, the
jet-dominated MAD models show less closure phase variability than the
disk-dominated SANE models, even in the large baseline triangles,
because the images from the latter are more sensitive to the
turbulence in the accretion flow. Our results suggest that image
reconstruction techniques need to explicitly take into account the
closure phase variability, especially if the quality and quantity of
data allow for a detailed characterization of the nature of variability.
This also implies that, if image reconstruction techniques that rely
on the assumption of a static image are utilized, regions of the $u-v$
space that show a high level of variability will need to be identified
and excised.

\end{abstract}


\section{Introduction}
The Event Horizon Telescope (EHT), a 1.3~mm wavelength VLBI experiment, will image, for the first time, black holes at event horizon scales (see, e.g., \citealt{2009astro2010S..68D}).
One of the primary observing targets for the EHT is Sagittarius~A$^*$ (Sgr~A$^*$), the supermassive black hole at the center of our galaxy.
Sgr~A$^*$ is an ideal candidate for the EHT since it has the largest angular size among the known nearby black holes \citep{2012ApJ...758...30J}, a well measured mass and distance  \citep{2008ApJ...689.1044G, 2009ApJ...692.1075G}, and has been extensively studied at a variety of wavelengths for over a decade (see \citealt{2001Natur.413...45B} and \citealt{2003Natur.425..934G} for early studies). 

The EHT will in principle measure visibility amplitudes and phases, which are the complex components of the Fourier transform of the image.
However, mm wavelength VLBI interferometers cannot measure absolute phases at each $u-v$ point covered by the array. 
This is because there are no point sources that are both close enough to Sgr~A$^*$ and bright enough at 1.3~mm to be used for calibration and because the timescale for variability of the atmospheric interference at 1.3~mm due to water vapor is only of the order of 10 s \citep{2002evn..conf..223D}. 
Instead, the EHT will measure closure phases, which are the sum of phases at three points in $u-v$ space, such that the effect of the atmosphere at each telescope cancels out \citep{1958MNRAS.118..276J}.
The EHT has already obtained closure phase data for Sgr~A$^*$ for the Hawaii, Arizona, California (HI-AZ-CA) triangle. \citet{2016ApJ...820...90F} reported a median closure phase of $+6.3^{\circ}$ over 13 observing nights during a four year period. 
The positive, non-zero average closure phase demonstrates that Sgr~A$^*$ is not circularly symmetric on event-horizon scales.

Even though closure phase measurements eliminate the variability due to atmospheric interference, they do not mitigate the effects of intrinsic source variability. 
Indeed, the flux from Sgr~A$^*$ has been observed to be variable at many wavelengths, including at 1.3~mm \citep[e.g.,][]{2008ApJ...682..373M, 2008A&A...488..549P, 2009ApJ...691.1021D}. The EHT observed variability at 1.3~mm on scales of a few Schwarzschild radii \citep{2011ApJ...727L..36F}.
The dynamical timescale, for Sgr~A$^*$ at event horizon scales is about ten minutes.
In contrast, the imaging timescale for the EHT is of the order of hours, because the interferometer relies on the rotation of the Earth to map out the $u-v$ space.
This points to the necessity of taking the intrinsic source variability into account when analyzing EHT data (see e.g., \citealt{2016ApJ...817..173L}) but also offers the potential of using source variability to probe the spacetime of the black hole near its horizon \citep{2009astro2010S..68D}.

A number of groups have considered the effects of closure phase variability in interpreting
EHT data. 
\citet{2009ApJ...695...59D} used a semi-analytic model to explore the variability
in closure phases caused by an orbiting hot spot for a few EHT closure triangles. 
\citet{2010ApJ...717.1092D} performed an early study of the properties of closure phase variability in GRMHD 
simulations focusing on disk dominated models and triangles which are appropriate for the already existing EHT observations.
\citet{2011ApJ...738...38B} compared stationary semi-analytic models with variable normalization to early EHT closure phase data.
\citet{2016ApJ...820..137B} studied closure phases for the HI-AZ-CA triangle in a stationary semi-analytic
accretion flow model, when small scale Gaussian brightness fluctuations were introduced.
\citet{2016A&A...588A..57F} used two GRMHD models, one jet dominated and one
disk dominated, to explore the effect of the Earth's rotation on the variability of closure phases in the
HI-AZ-CA triangle but did not include the effect of intrinsic source variability.

In this paper, we aim to characterize the expected properties of closure phases for Sgr~A$^*$ in
a wide range of EHT triangles of various sizes and orientations, using a suite of disk- and jet-dominated 
GRMHD+radiative transfer simulations.
 We employ five long time span models that probe a range of 
black hole spins, initial magnetic field  geometries, and thermodynamic prescriptions for the 
electrons (see \citealt{chan2015, 2015ApJ...799....1C}
for the details of the models). The parameters of the models, which we review in \S2, have been calibrated to fit the
broadband spectra, the 1.3 mm image size, and the multi wavelength variability of Sgr~A$^*$. In
\S3, we investigate the expected magnitudes of the interferometric visibility phases throughout
the $u-v$ plane. Even though the Event Horizon Telescope will not be able to measure directly
the visibility phases at individual locations on the $u-v$ plane, exploring their properties allows
us to understand in \S4 the variability of the closure phases that the Event Horizon Telescope
will measure. We conclude in \S5 and compare our results to the existing limited number of
closure phase measurements from Sgr~A$^*$ on a single baseline triangle \citep{2016ApJ...820...90F}.

\section{The GRMHD+ray Tracing Simulations}\label{sec:sims}

In previous work, we considered a large number of GRMHD + radiative transfer simulations where we varied the black hole spin, the initial geometry of the magnetic field, the accretion rate, and the thermodynamic prescription for the electrons \citep{2015ApJ...799....1C}. 
We created these models using \texttt{HARM} \citep{2003ApJ...589..444G} for the GRMHD simulations  \citep{sadowski_fix, 2013MNRAS.436.3856S} and \texttt{GRay} \citep{2013ApJ...777...13C} to solve the radiative transfer equation along null geodesics. 
As discussed in \citet{2013ApJ...777...13C}, our dynamical simulations were performed under the assumptions of ideal MHD and radiative transfer calculations were performed using the fast-light approximation.
We calibrated the simulations using Sgr~A$^*$ data. 
Specifically, we enforced the following constraints:
{\em (a)\/} a flux and a slope in the $10^{11}$--$10^{12}$ Hz range that matches observations, 
{\em (b)\/} a flux at $\simeq 10^{14}$ Hz that falls within the observed range of the highly variable infrared flux, 
{\em (c)\/} an X-ray flux that is consistent with $10\%$ of the observed quiescent flux, i.e., the percentage which has been attributed to emission from the inner accretion flow \citep{2013ApJ...774...42N}, and 
{\em (d)\/} a size of the emission region that is consistent with the size determined by the early EHT observations \citep{2008Natur.455...78D}. 

We identified 5 models that fit all observational constraints. 
All models have an observer inclination of $i = 60^{\circ}$ with respect to the spin axis of the black hole.
Model A has a black hole spin of $a=0.7$, an initial magnetic field geometry that leads to weak, turbulent fields near the horizon
(SANE, Standard and Normal Evolution), a constant electron-ion temperature ratio for the thick accretion disk, and a constant electron temperature for the funnel region. 
Model B is the same as Model A but with a black hole spin of $a=0.9$. 

Models C-E, in contrast, have an initial magnetic field geometry that leads to coherent magnetic field structures near the horizon (MAD, Magnetically Arrested Disk).
In addition, Model C has a black hole spin of $a=0.0$, a constant electron-ion temperature ratio for the disk and a different electron-ion temperature ratio for the funnel region. Model D has a black hole spin of $a=0.9$ and, again, a constant electron-ion temperature ratio for the disk and a constant electron temperature for the funnel region. Model E is like Model D but with a constant electron-ion temperature ratio for the funnel region. 
For all models, the values for the normalization of the electron number density,  the electron-ion temperature ratio for the disk and funnel and/or the electron temperature for the funnel were fit to the observations described above (see \citealt{2015ApJ...799....1C} for detailed model parameters).  
We use 1024 snapshots from each simulation with a time resolution of $10GMc^{-3}$ or $\approx$3.5 minutes, which results in a total time span of approximately 60 hours. Table \ref{tab:fit} summarizes the parameters of the models we consider.

\begin{deluxetable}{ccllrr}
\tablewidth{\columnwidth}
\tablecaption{Summary of our Five Models}  
\tablehead{Name & $a$ & \ \ $B_0$ & \ Plasma Model \ }
\startdata
A & 0.7 & SANE & Constant $\Tf$ \\
B & 0.9 & SANE & Constant $\Tf$  \\
C & 0.0 & MAD  & Constant $\tf$  \\
D & 0.9 & MAD  & Constant $\Tf$\\
E & 0.9 & MAD  & Constant $\tf$ 
\enddata
\tablecomments{Summary of the five best fit models from \citet{2015ApJ...799....1C}. 
The first column lists the model names used throughout the paper. 
The second and third columns list the black hole spin ($a$) and the accretion flow state that depends on the initial magnetic field geometry ($B_0$).
The fourth column refers to the plasma model used in the funnel region, specifically $\Tf$ refers to a constant electron temperature while $\tf$ refers to a constant electron-to-ion temperature ratio. }
\label{tab:fit}
\end{deluxetable}

\begin{figure*}[t!]
\centering
\includegraphics[height=2in]{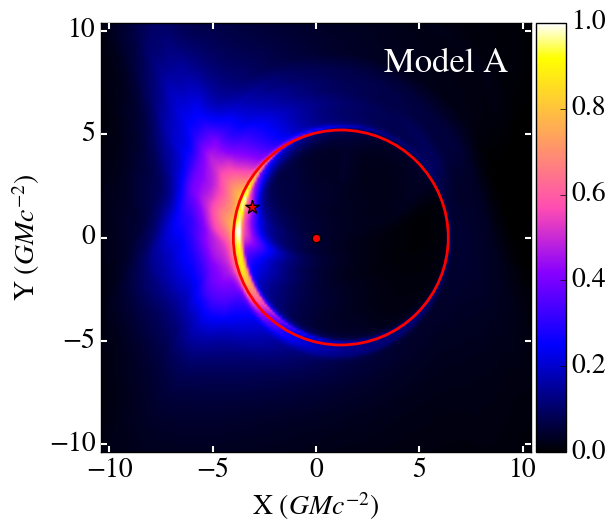}
\includegraphics[height=2in]{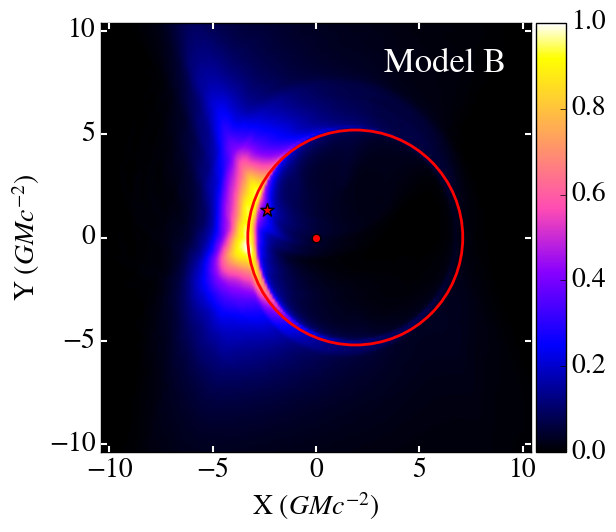}
\includegraphics[height=2in]{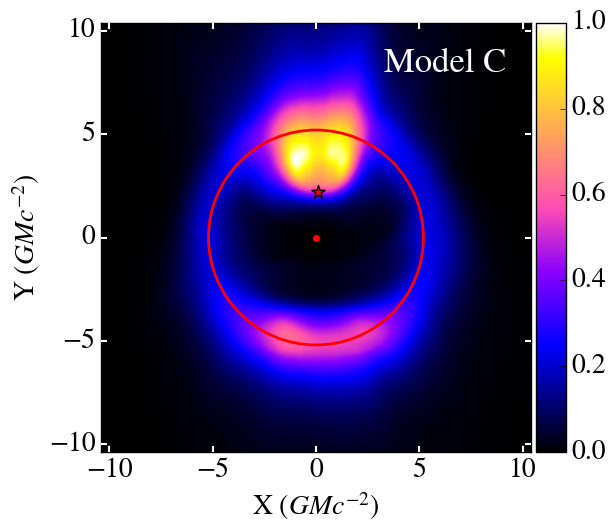}
\includegraphics[height=2in]{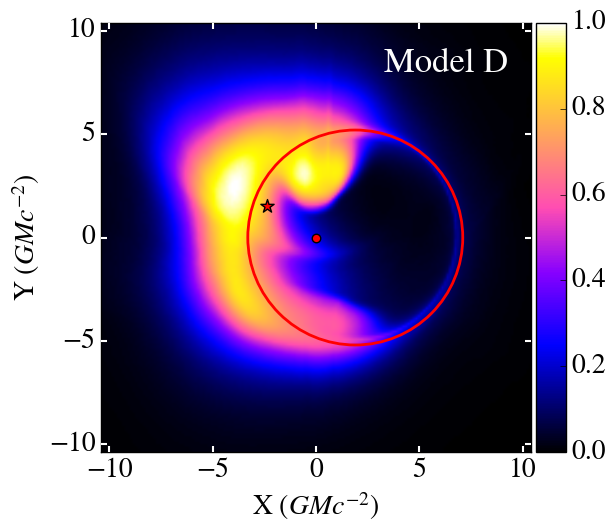}
\includegraphics[height=2in]{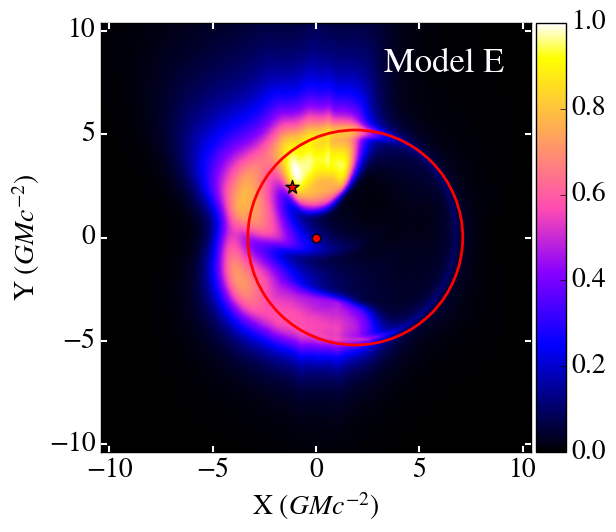}
\caption{The average 1.3~mm images of the five models we consider in this paper. The SANE models (A and B) have most of their emission originating from the disk region, while the MAD models (C, D, and E) have significant emission originating from the jets. Model C is unique, with negligible emission from the disk and a black hole spin of zero.  The red circles indicate the expected size of the black hole shadow according to general relativity. The red stars correspond to the location of the center of light for each model while the red dots are the location of the center of the black hole. Since the orientation of Sgr~A$^*$ on the sky is not known, these images show an arbitrary orientation where the spin axis of the black hole points North. The maximum intensity in each panel has been normalized to unity.}
\label{fig:averageS}
\end{figure*}

Figure \ref{fig:averageS} shows the average 1.3~mm wavelength images for the five simulations we consider. 
Models A and B, the SANE models, have 1.3~mm wavelength emission regions that are dominated by the thick accretion disk.  
The emission from these models is asymmetric due to the effects of relativistic Doppler beaming, because the part of the orbiting accretion flow that is coming towards the observer is beamed and appears brighter than the part that is moving away from the observer. 
The MAD models, (C, D, and E), however, have 1.3~mm wavelength emission regions that are dominated by the funnel regions. 
Model C is unique in that its emission is dominated by the footprints of the outflows with negligible emission coming from the disk.
Models D and E have emission coming from both the Doppler beamed disk and the outflows.

The red circles superimposed on the images correspond to the size of the black hole shadow predicted by general relativity. 
For comparison, the red dots show the location of the center of the black hole.
Frame dragging effects cause the emission to be offset (red circles are not centered on the black hole) for models which have a non-zero spin.
The red stars in the figures indicate the calculated center of light for each image. 
The center of light was used in the calculation of the Fourier transform of the image, as we will  discuss below.

\section{Visibility Phases}\label{sec:phases}

\begin{figure*}[t!]
\centering
\includegraphics[height=1.95in]{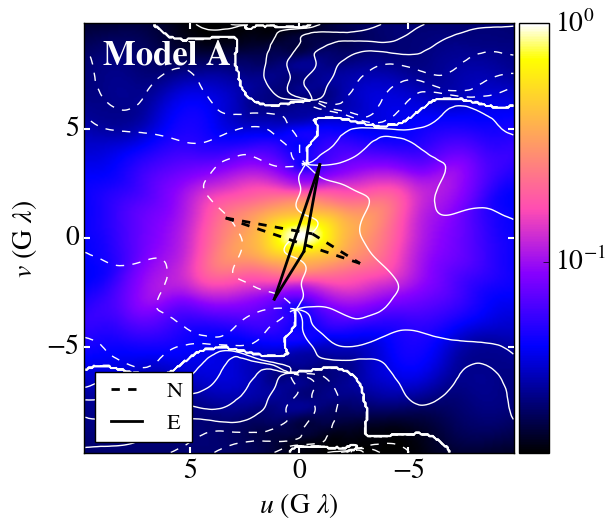}
\includegraphics[height=1.95in]{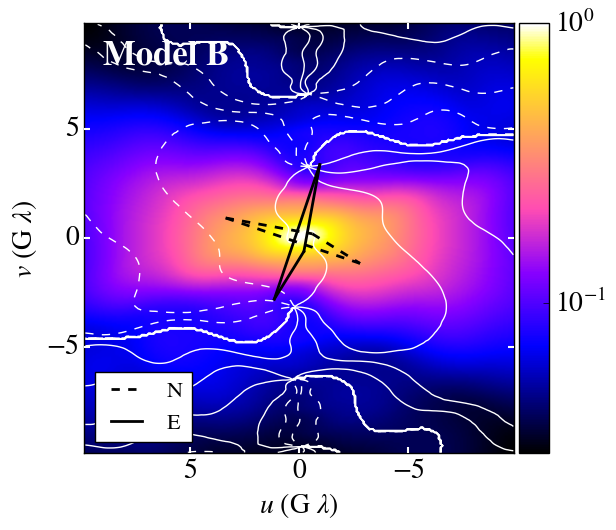}
\includegraphics[height=1.95in]{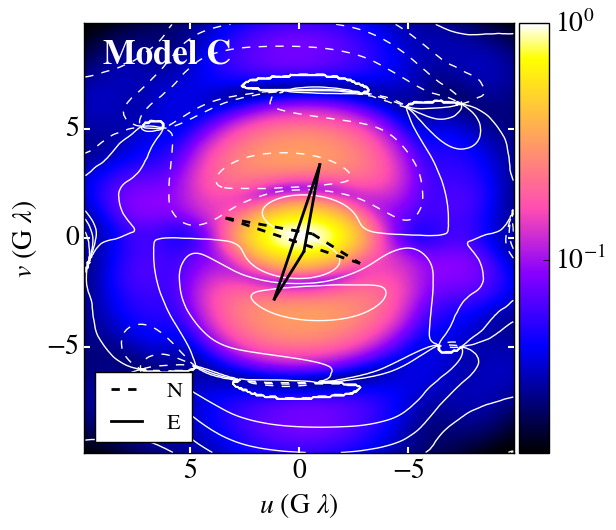}
\includegraphics[height=1.95in]{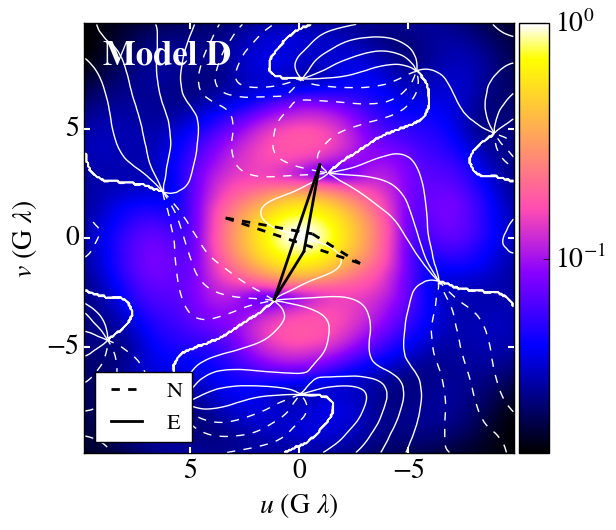}
\includegraphics[height=1.95in]{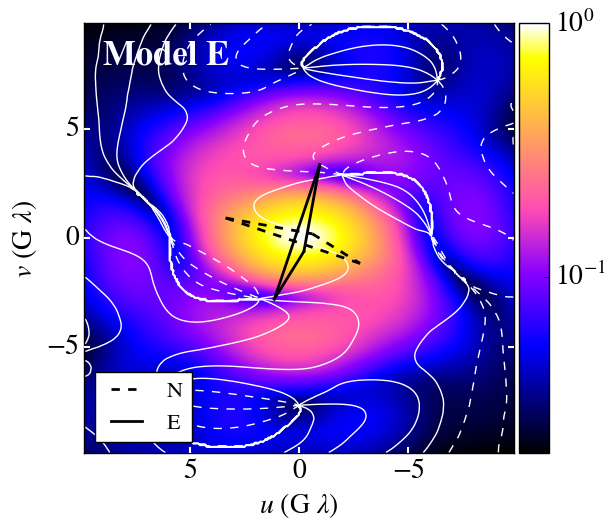}
\caption{The color maps show the average visibility amplitudes and the white contours the average visibility phases for the five models. Minima in the visibility amplitudes correspond to steep gradients in visibility phases. The black dashed triangles correspond to the HI-AZ-CA triangle for a black hole with a spin axis pointing North while the solid black triangles correspond to the same HI-AZ-CA triangle for a black hole with spin axis pointing East. The visibility amplitude maps have been normalized to unity.
\\
\\}
\label{fig:VA_VP_contour}
\end{figure*}

\begin{figure*}[t!]
\centering
\includegraphics[height=1.95in]{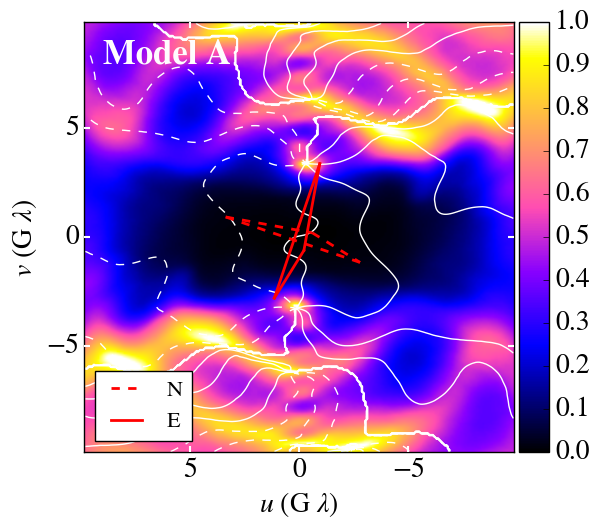}
\includegraphics[height=1.95in]{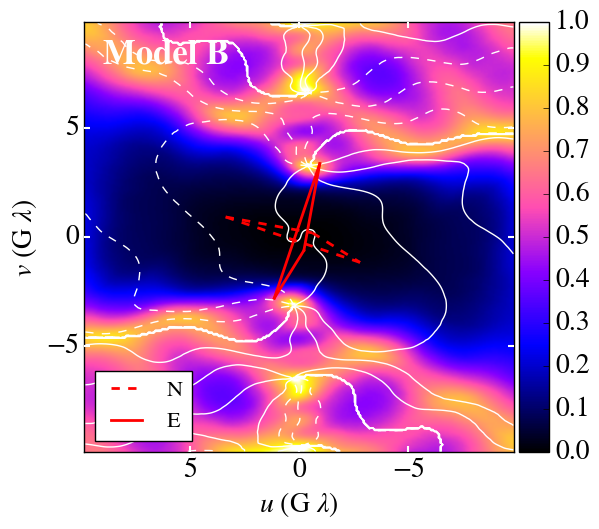}
\includegraphics[height=1.95in]{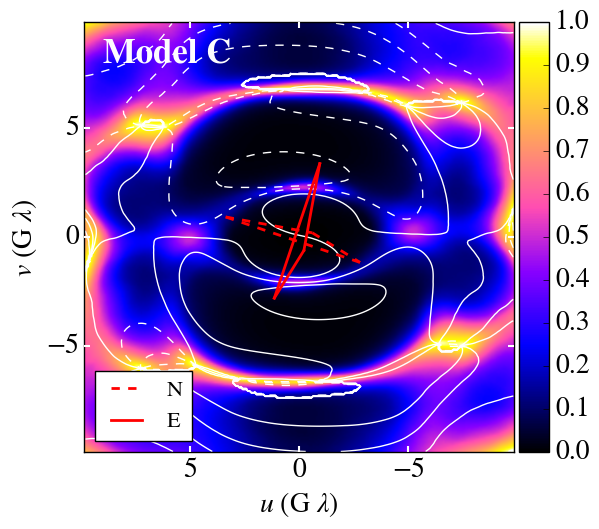}
\includegraphics[height=1.95in]{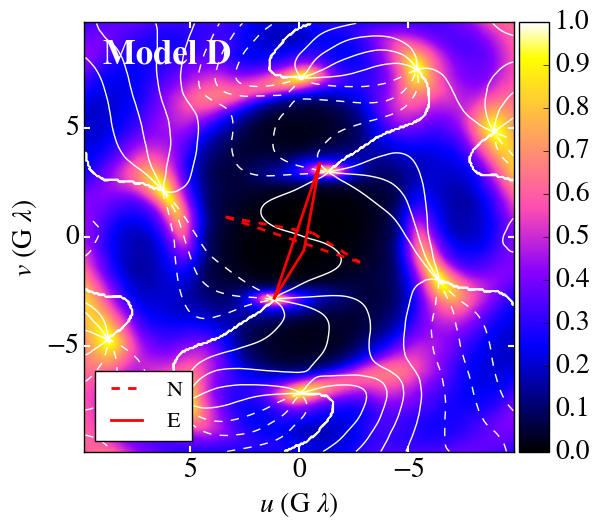}
\includegraphics[height=1.95in]{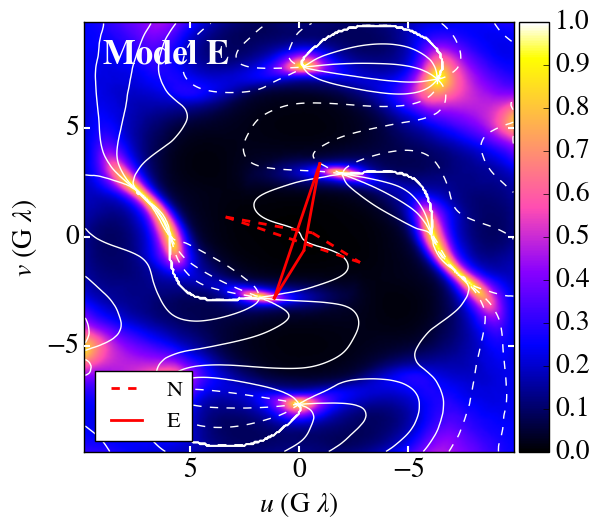}
\caption{The color map shows the dispersion in visibility phase at each point in the $u-v$ space throughout each $\sim60$~hour simulation. These dispersions were calculated using the directional statistics described in the Appendix. The white contours correspond to the average phase. Regions of steep phase gradients (and minimum amplitudes; cf. Figure \ref{fig:VA_VP_contour}) correspond to large dispersion in visibility phase. The red dashed (solid) triangle corresponds to the HI-AZ-CA closure triangle for a black hole with spin axis pointing North (East).
\\
\\}
\label{fig:RMS_contour}
\end{figure*}

Since the EHT is an interferometer, it will observe the visibilities, or the complex Fourier components, of the image of Sgr~A$^*$.
The amplitudes of these Fourier components, or visibility amplitudes, for our five models have been discussed in \citet{2016arXiv160106799M}. 
Here we focus on the phases of the complex Fourier components, or visibility phases. 

Due to the effects of gravitational lensing and Doppler beaming, the emission predicted by these models is not centered on the black hole (the red stars and red dots in Figure~\ref{fig:averageS} are in different locations), which results in an overall rapid gradient in phase. 
We removed this unmeasurable phase gradient by shifting the snapshots such that the center of light of the images (the red stars in Figure~\ref{fig:averageS}) coincide with the center of the average image (red dots in Figure~\ref{fig:averageS}) before calculating the transforms. 
We performed the same shift for all snapshots within each simulation such that they all have the same phase centers. 

\begin{figure*}[t!]
\centering
\includegraphics[width=\textwidth]{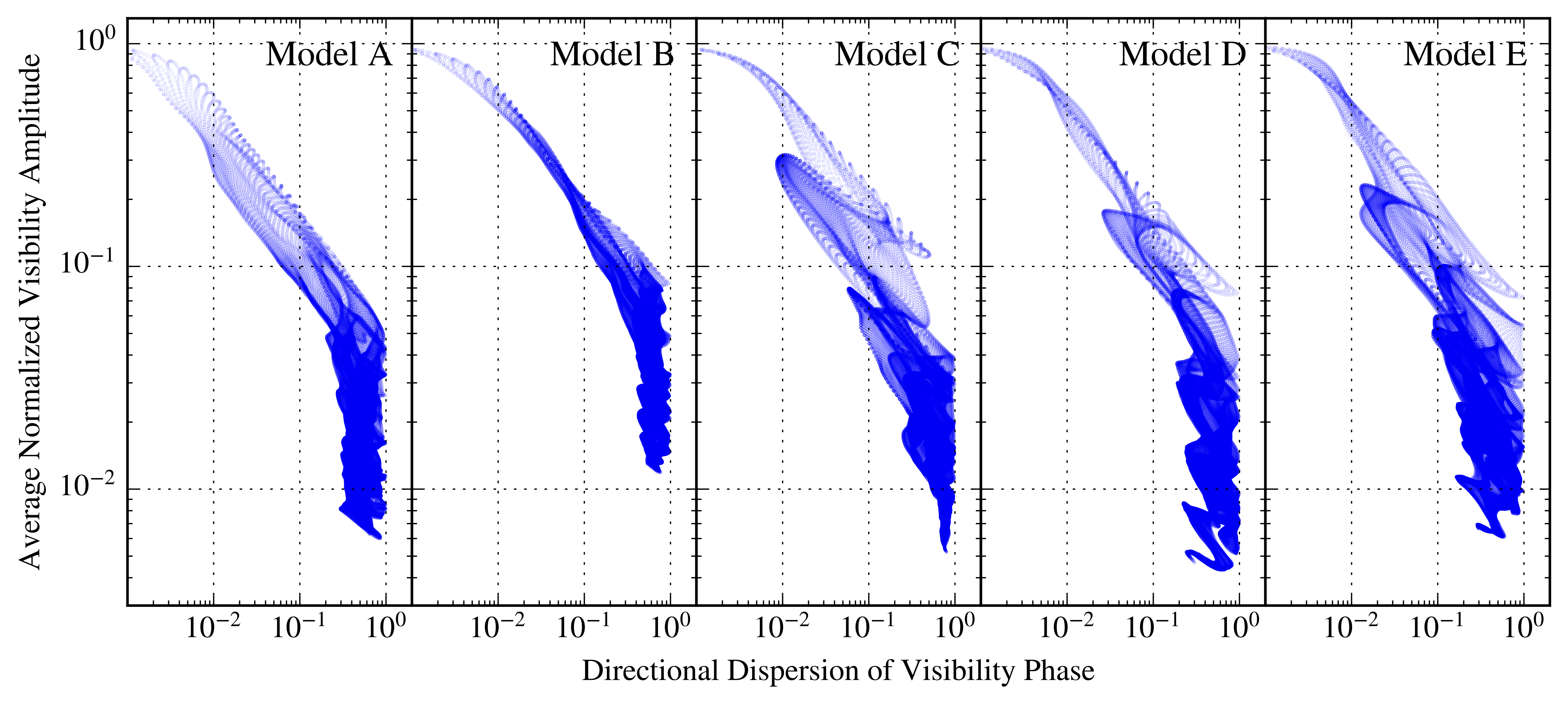}
\caption{Normalized visibility amplitude versus directional dispersion ($D$) of visibility phase. 
The different columns correspond to the different models and each dot is a point in $u-v$ space. In all models, the regions of largest dispersion in visibility phase correspond to the lowest visibility amplitudes.}
\label{fig:VA_rmsVP}
\end{figure*}

In Figure \ref{fig:VA_VP_contour}, we present the structure of the complex visibilities for the different GRMHD models denoting the average visibility phases with contours and the visibility amplitudes in color maps.
These averages are obtained by finding the phases and amplitudes of each snapshot and subsequently averaging them.
Because angles are directional, periodic quantities, we need to employ a method for calculating means that is appropriate for them.
In the Appendix, we describe the directional statistic we use hereafter. 

Figure \ref{fig:VA_VP_contour} highlights the fact that minima in visibility amplitudes coincide with steep gradients in phase. 
This is particularly prominent in the MAD models (C, D, and E), which have clear minima that are preserved in the average of the visibility amplitudes as we showed in our previous paper \citep{2016arXiv160106799M}.
In each panel, the black dashed (solid) triangle corresponds to the HI-AZ-CA closure triangle for a 
black hole with spin axis pointing North (East). 
For brevity, we plotted the visibility amplitude and phase averages of the black hole at a constant (North) orientation but moved the triangle so that the relative orientation of the visibilities and the triangles is correct for the quoted black hole spin axis orientation. 
In reality, the orientation of the triangle is fixed and the orientation of the black hole in the sky is unknown. 
We will discuss these triangles further below. 

We explore the structure of the variability in the visibility phases in Figure \ref{fig:RMS_contour}, where the dispersion in visibility phases is shown in color, while the average visibility phases are shown as white contours for comparison. 
The dispersion was calculated by taking into account the fact that angles are directional quantities, as discussed in the Appendix.
The directional dispersion, $D$ (see equation A4), is approximately equal to  $\sigma^2/2$ for small $\sigma$, where $\sigma$ is the dispersion of a non-directional Gaussian distribution, and approaches unity in the limit of a flat distribution. 
In the figure, the black and dark blue regions have small dispersions, while the yellow or white regions have very broad, and possibly flat distributions. 
For reference, $D=0.5$ shown in pink corresponds to a Gaussian with a standard deviation of about 1 rad or 57$^{\circ}$.

Figure \ref{fig:RMS_contour} shows two general characteristics of variability in
the visibility phase throughout the $u-v$ plane. First, for each
model, there are a number of localized regions on the $u-v$ plane that
exhibit very large dispersions in phase. These regions coincide with
the locations of the minima in the visibility amplitudes. 
Phase variability is related to visibility amplitude since, for similar perturbations in its real and imaginary components, a vector with larger magnitude will experience a smaller change in phase.
This means that, if all complex Fourier components of the image experience perturbations to their real and imaginary components of similar size, the complex vectors with smaller amplitude will experience a larger change in phase than those with a larger amplitude. 
Therefore the regions that have low visibility amplitude also have very high phase variability as a direct consequence of the low amplitude (see also the discussion in \citealt{2010ApJ...717.1092D}.)

Second, outside the confined locations of the amplitude minima, the
variability in the visibility phases at small baseline lengths (less
than a few G$\lambda$) is in general very small, even though the
accretion flow is highly turbulent. This happens because the small
baselines primarily probe the overall structure of the image, which is
determined by special and general relativistic effects rather than gas dynamics, and shows
little variability. However, at larger baseline lengths, for most
baseline orientations, the SANE models A and B show significant phase
variability, while the MAD models C-E remain relatively quiet. The
large baselines probe the small scale structures, which, in the case
of the SANE models, are dominated by, e.g., small hot magnetic flux
tubes that are highly variable. In the case of the MAD models, even
the small scale structure is dominated by the emission at the jet
footpoints and, for the models we consider here, the closure phases along large baseline triangles are not significantly
variable.

In order to demonstrate one of the above points in a different way, we
show in Figure \ref{fig:VA_rmsVP} the overall anticorrelation between the average
visibility amplitude throughout the $u-v$ plane and the corresponding
dispersion in the visibility phase. Indeed, the largest phase
dispersions occur when the visibility amplitude is very low, i.e., at
least an order of magnitude smaller than its maximum. As this figure demonstrates, this
anticorrelation is independent of the particular cause of variability
or the specifics of the models explored.

\begin{figure}[t!]
\centering
\includegraphics[height=3.4in]{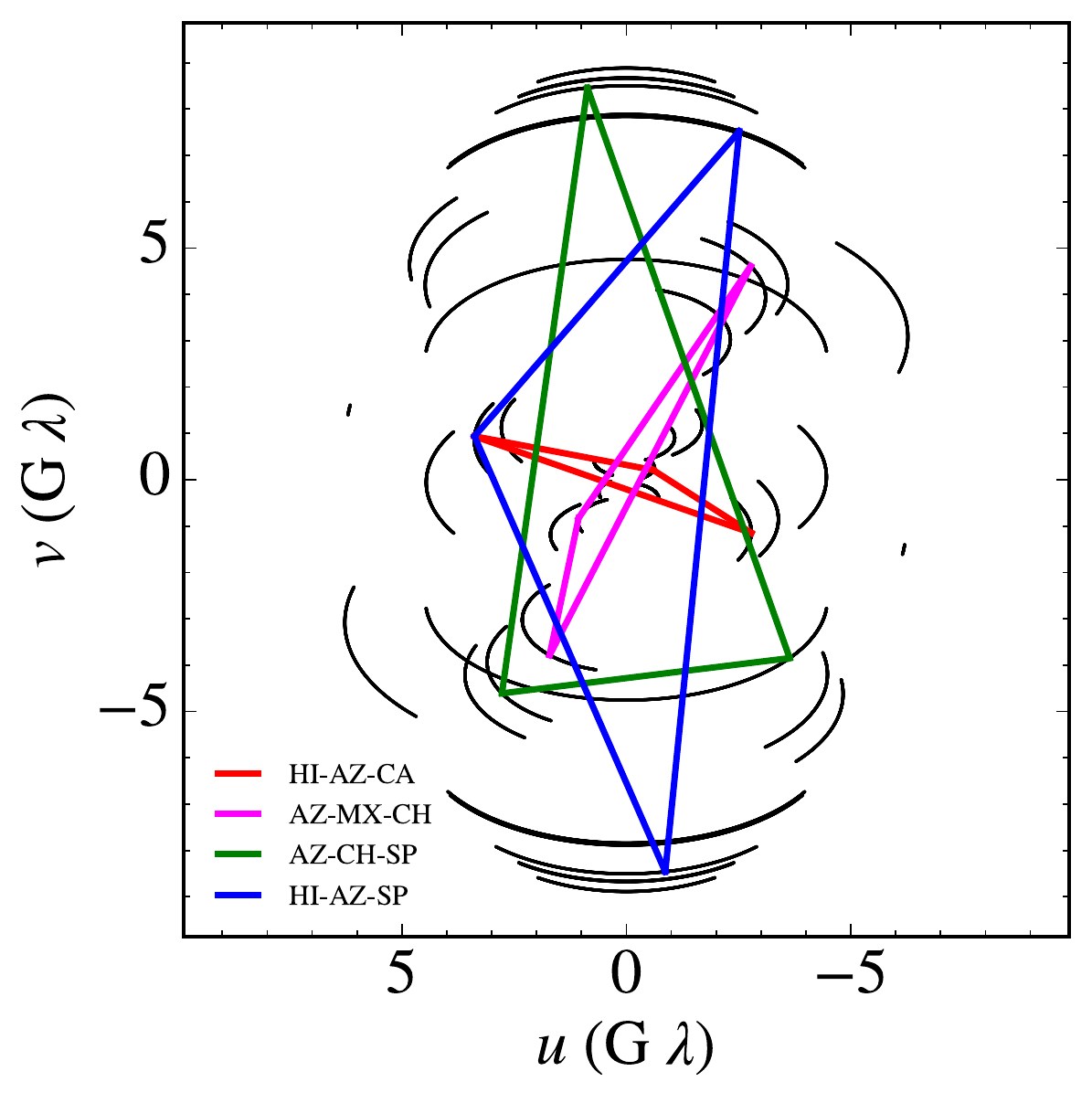}
\caption{The four triangles we used to calculate closure phases (shown here at GMST 01:54:03.4706). In order of increasing size, they are Hawaii (SMA)-Arizona (SMT)-California (CARMA), shown in red, Arizona (SMT)-Mexico (LMT)-Chile (ALMA), shown in magenta, Arizona (SMT)-Chile (ALMA)-South Pole (SPT), shown in green, and Hawaii (SMA)-Arizona (SMT)-South Pole (SPT), shown in blue. The black curves correspond to the EHT baselines for reference. These triangles move through $u-v$ space following the baseline tracks as the Earth rotates. }
\label{fig:triangles}
\end{figure}

\begin{figure*}[p!]
\centering
\includegraphics[height=3.in]{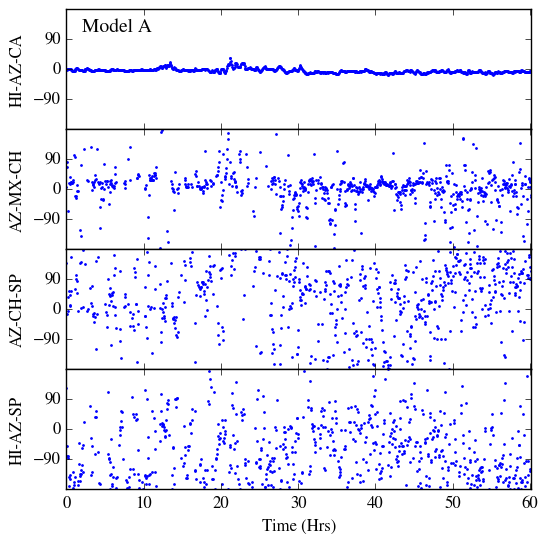}
\includegraphics[height=3.in]{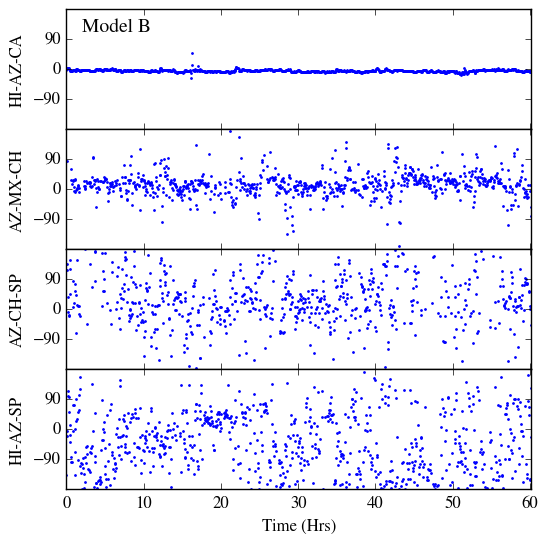}
\includegraphics[height=3.in]{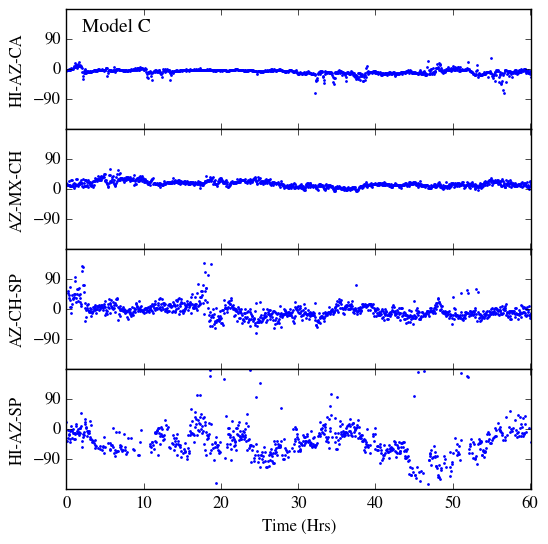}
\includegraphics[height=3.in]{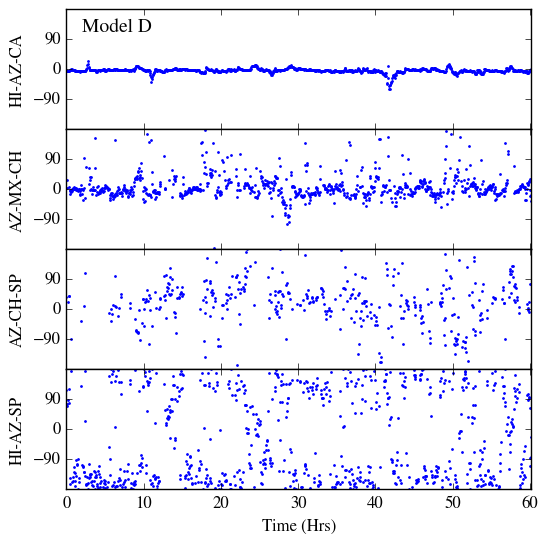}
\includegraphics[height=3.in]{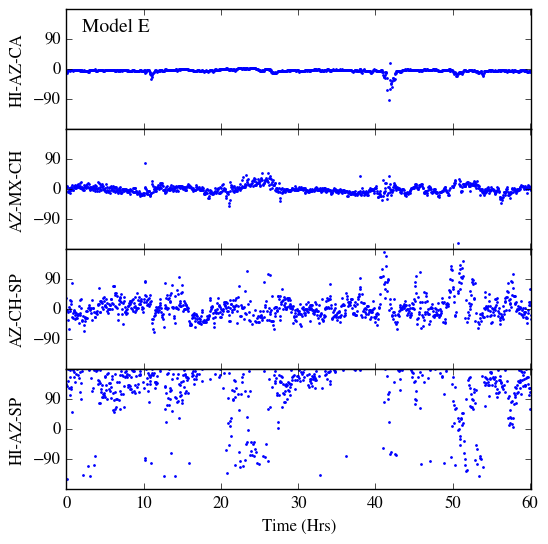}
\caption{Closure phases as a function of time for the five simulations and the four closure triangles we consider for a black hole with a spin axis pointing North. Different rows correspond to different triangles in order of increasing size from top to bottom. 
}
\label{fig:CPvTN}
\end{figure*}
\begin{figure*}[p!]
\centering
\includegraphics[height=3.in]{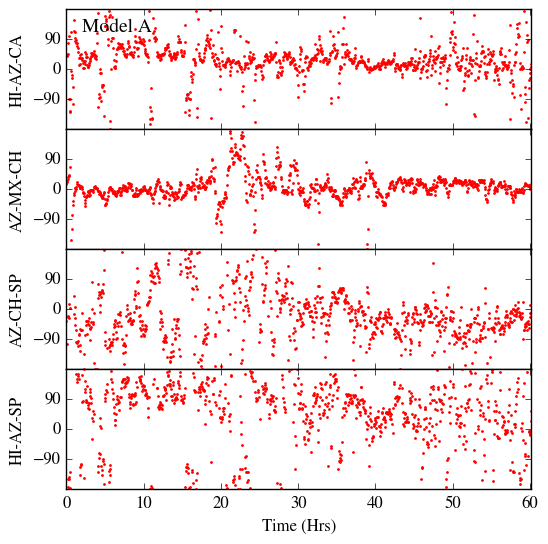}
\includegraphics[height=3.in]{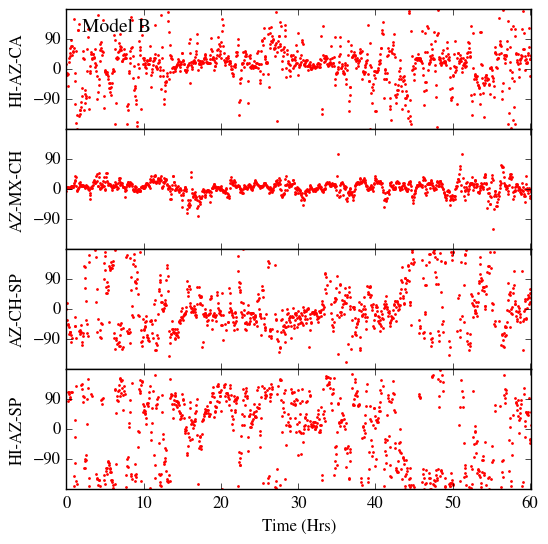}
\includegraphics[height=3.in]{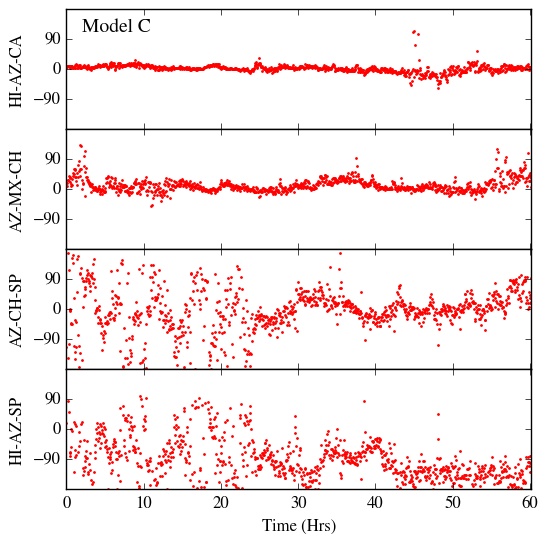}
\includegraphics[height=3.in]{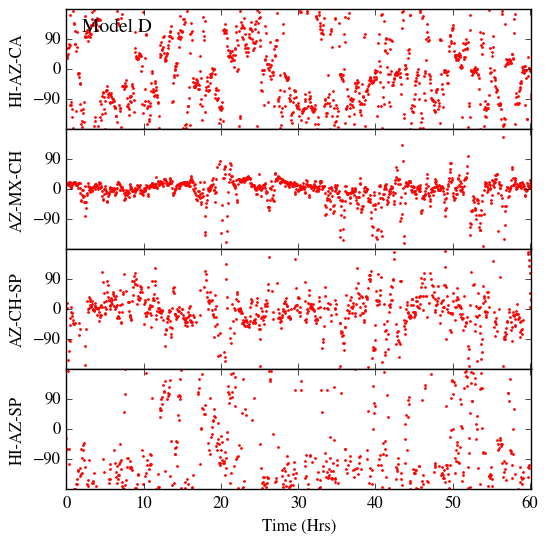}
\includegraphics[height=3.in]{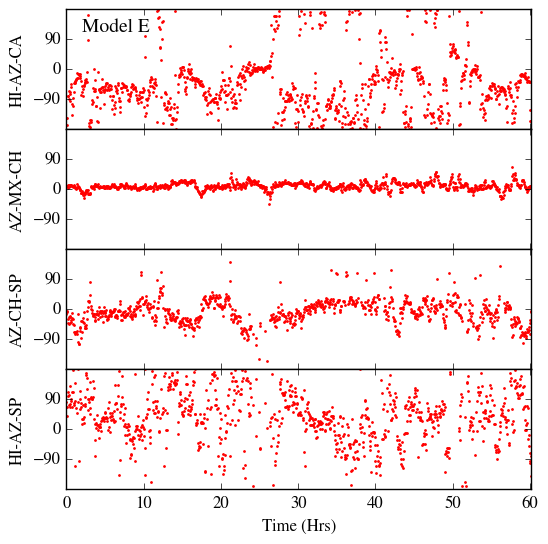}
\caption{Same as Figure \ref{fig:CPvTN} but for a black hole with spin axis pointing East. 
}
\label{fig:CPvTE}
\end{figure*}

\begin{figure*}[p!]
\centering
\includegraphics[width=6in]{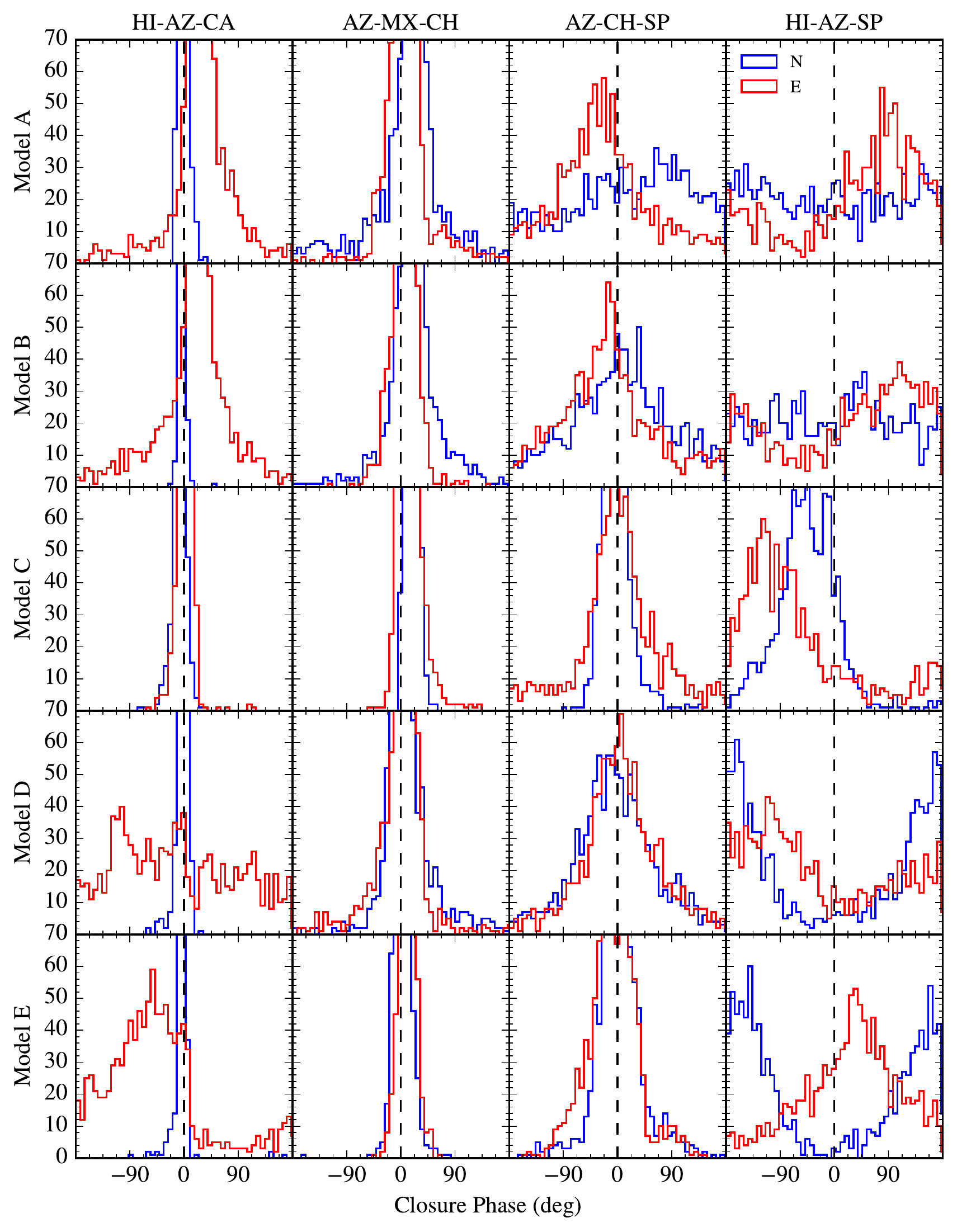}
\caption{Histograms of closure phases calculated for the four triangles we consider. Rows correspond to the different models and columns to different closure triangles, in order of increasing size. In all panels, the blue histograms correspond to a black hole with spin axis that points North, while the red histograms correspond to a spin axis that points East. The orientation of the black hole has a very large effect on the mean and width of the distribution of the closure phases. Some of the peaks of the histograms are not shown in this figure since we are not concerned primarily with the value of the peak but rather with the width of the distribution.}
\label{fig:hist}
\end{figure*}

\section{Closure Phases} \label{sec:CP}
As we discussed in the previous sections, mm VLBI experiments cannot measure absolute phase since the atmosphere introduces an arbitrary phase that is variable on a $\approx10$ s timescale. 
Instead, the EHT measures closure phases, defined as the sum of the phases at the corners of a triangle in $u-v$ space that corresponds to three telescopes on Earth. 
Measuring closure phases removes the effects of the atmosphere and instrumental noise from the phase measurements, 
but cannot recover all absolute phase information, because there are never enough closure triangles to solve for all absolute phases. 

Our aim is to explore what the closure phases that the EHT measures will reveal about horizon-scale structures and how they will probe small-scale variability.
During the span of an observation, closure triangles move through the $u-v$ space. 
Therefore, the observed variability in closure phases will reflect the combined effect of the intrinsic variability of the source, the variability caused by the fact that the closure triangles are probing different parts of $u-v$ space as the Earth rotates (see, e.g., \citealt{2009ApJ...695...59D,2011ApJ...738...38B,2016ApJ...820..137B,2016A&A...588A..57F}), and the variability caused by diffractive scattering effects (see \citealt{2015ApJ...805..180J}). 
In the current section, we are primarily interested in exploring the intrinsic variability caused by the accretion flow itself. Because of this, we keep the triangles constant in time for the majority of our analysis (fixed at GMST 01:54:03.4706); we will explore the effect of the Earth's rotation at the end of this section and the effects of scattering in a forthcoming paper.  

We choose 4 representative triangles of varying shapes and sizes, shown in Figure \ref{fig:triangles}.
The smallest triangle, shown in red, is the Hawaii (Submillimeter Array-SMA)-Arizona (Submillimeter Telescope-SMT)-California (Combined Array for Research in Millimeter-wave Astronomy-CARMA) triangle. 
This is the only triangle on which the EHT has observed closure phases to date.
The next smallest triangle, shown in magenta, is the Arizona (SMT)-Mexico (Large Millimeter Telescope-LMT)-Chile (Atacama Large Millimeter/submillimeter Array-ALMA) triangle. 
The bigger triangles are Arizona (SMT)-Chile (ALMA)-South Pole (South Pole Telescope-SPT), shown in green, and Hawaii (SMA)-Arizona (SMT)-South Pole (SPT), shown in blue. 

We calculate closure phases for the four triangles for each
snapshot of our five models, for black holes with spin axes pointing
North and East. Due to the symmetry of Fourier transforms, the closure
phases for black holes with spin axes pointing South (West) are the
negative of the closure phase for black holes pointing North
(East). We use the same sign convention as described in \citet{2016ApJ...820...90F}.
Figures \ref{fig:CPvTN} and \ref{fig:CPvTE} show closure phases as a function of time for both
spin orientations, for the four triangles (ordered from the smaller triangle on the top
row to the largest on the bottom row), and for the five models.  To explore
the distribution of closure phases more quantitatively, we also plot
them as histograms in Figure \ref{fig:hist}.

As the top panels for each model in Figures~\ref{fig:CPvTN} and \ref{fig:CPvTE} and the leftmost panels
in Figure \ref{fig:hist} show, all of our GRMHD models produce little phase
variability on small triangles, with the exception of situations where
at least one vertex of the triangle crosses an amplitude minimum (see,
e.g., the East orientation for models D and E).  On larger triangles,
the closure phases generally show larger dispersion. For these
triangles, however, there is an important difference between the SANE
and the MAD models. The MAD models still show peaked distributions of
closure phases with well defined means and dispersions, whereas the
histograms of the SANE models become nearly flat. Both of these
results are expected given our discussion of visibility phase
variability in Section 3.

\begin{figure}[t!]
\centering
\includegraphics[width=3.5in]{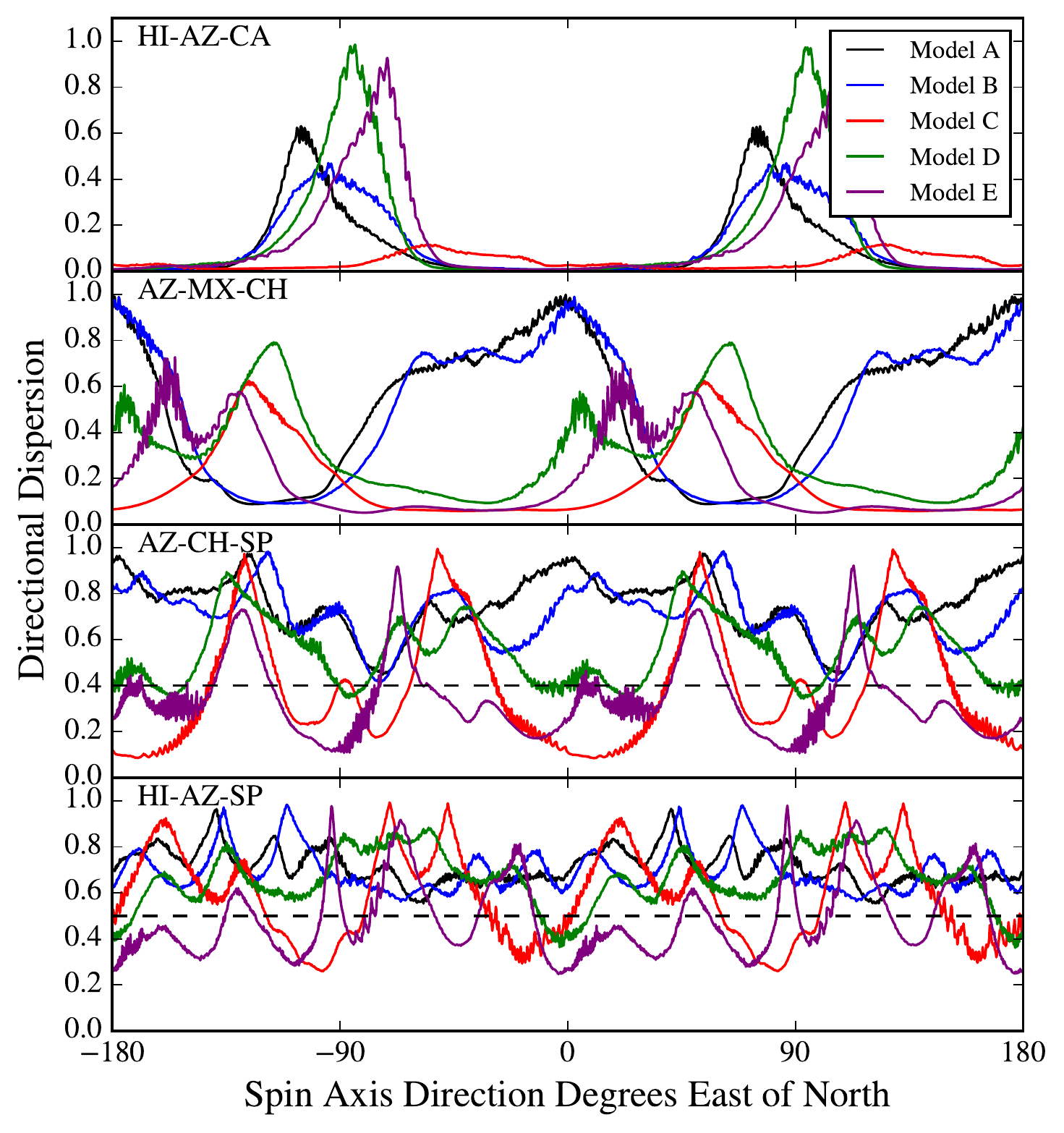}
\caption{ Directional dispersion of closure phase for the four closure phase triangles we consider as a function of the orientation of the black hole spin axis.
The black dashed line in the third panel corresponds to a directional dispersion of 0.4 which corresponds to a gaussian with a dispersion of about $51^{\circ}$.
The black dashed line in the fourth panel corresponds to a directional dispersion of 0.5 which corresponds to a gaussian with a dispersion of about $57^{\circ}$.}
\label{fig:orient}
\end{figure}

The results shown in Figures~\ref{fig:CPvTN}, \ref{fig:CPvTE} and \ref{fig:hist} are not specific to the
particular black-hole spin orientations chosen for these examples but
are generically encountered in all orientations.  We demonstrate this
in Figure~\ref{fig:orient}, which shows the dependence of the closure phase
dispersions on black-hole orientation, for the four triangles and for
the five models we consider here. In the smallest of the triangles, a
small phase dispersion (at the level recently reported by \citealt{2016ApJ...820...90F}) 
occurs for about half of the spin-orientation parameter space for all
five models. However, for the largest triangles, the large dispersion
in the SANE models persists for all spin orientations.

The statistical properties of closure-phase variability that we
discussed so far correspond to fixed orientations of the baseline
triangles on the $u-v$ plane. In practice, we can observationally
infer these properties if we combine data from different epochs and
stack them based on the location of each triangle on the $u-v$ plane.
However, in the course of a single observation epoch, the orientation
of each baseline triangle changes in time and the measured closure
phases will sample different locations of the $u-v$ plane, while the
underlying image is varying at the same time. A consequence of this
may be that a given triangle will rotate from a region of small
variability to one of large variability (e.g., near a visibility
minimum) or vice versa in the course of a night. In this case, the
characteristics of phase variability will change dramatically in the
course of the observation.

\begin{figure*}[t!]
\centering
\includegraphics[height=2in]{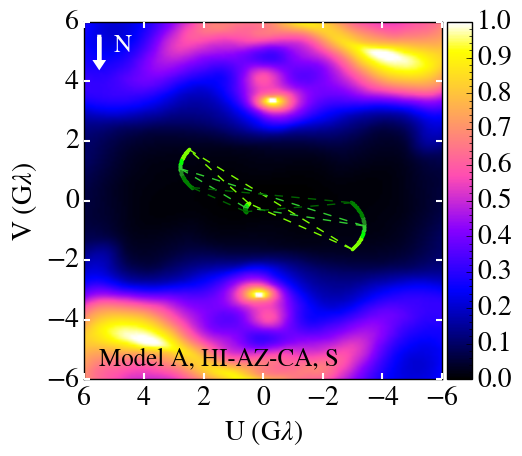}
\includegraphics[height=2in]{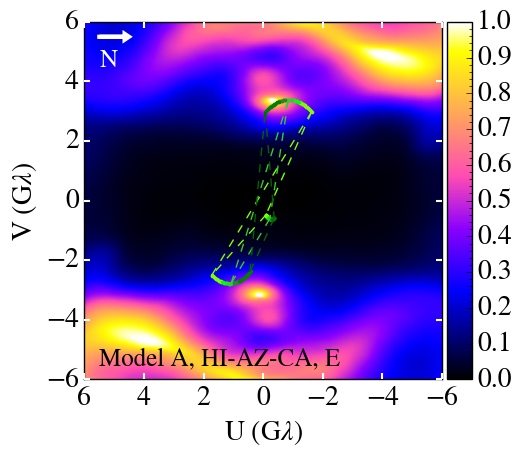}
\includegraphics[height=2in]{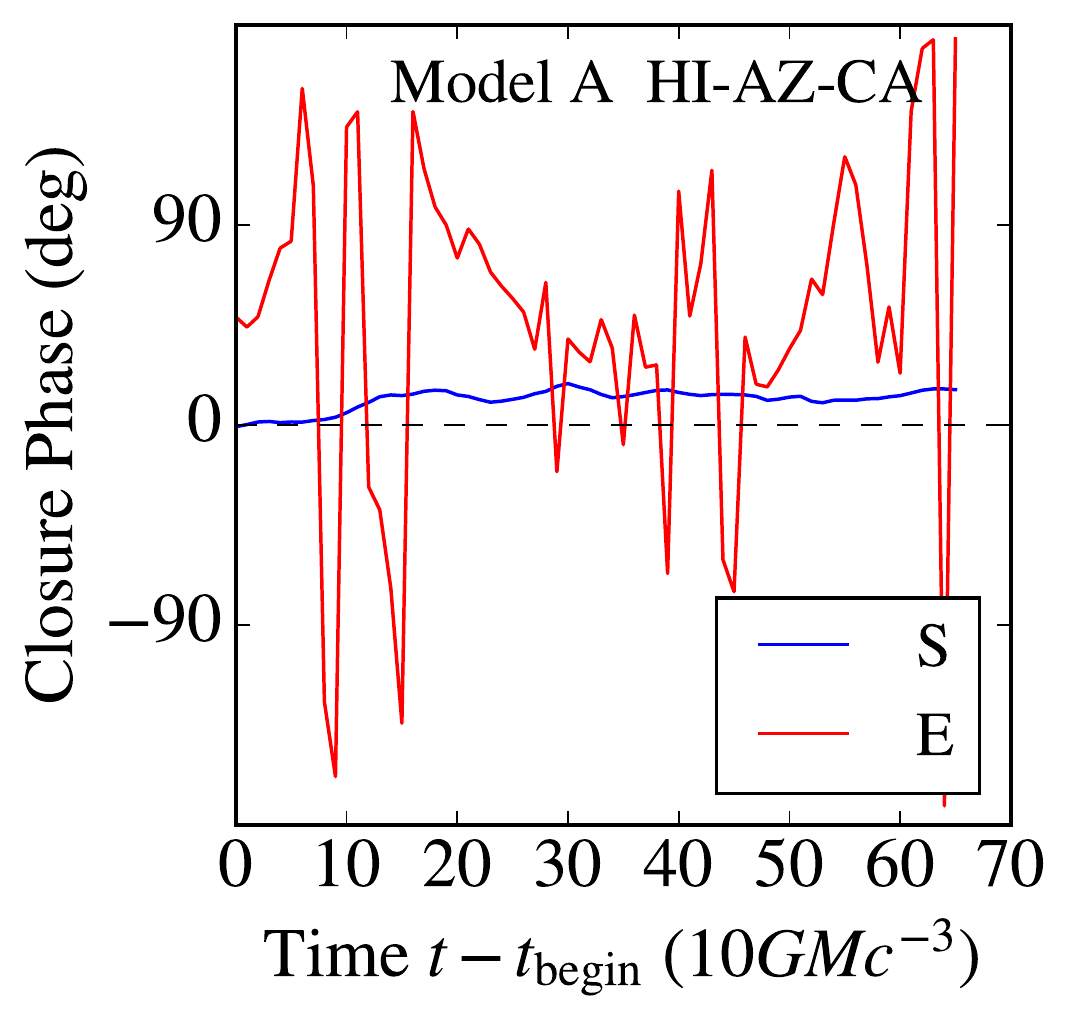}
\includegraphics[height=2in]{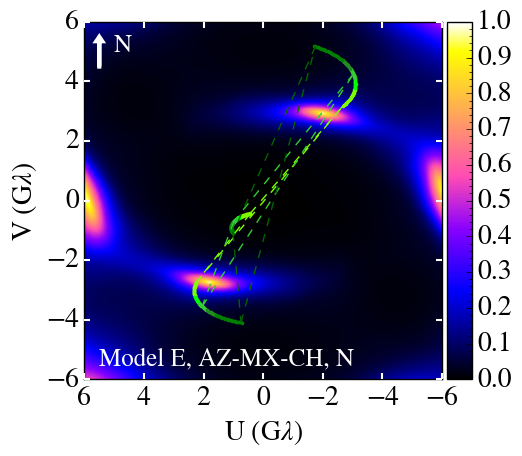}
\includegraphics[height=2in]{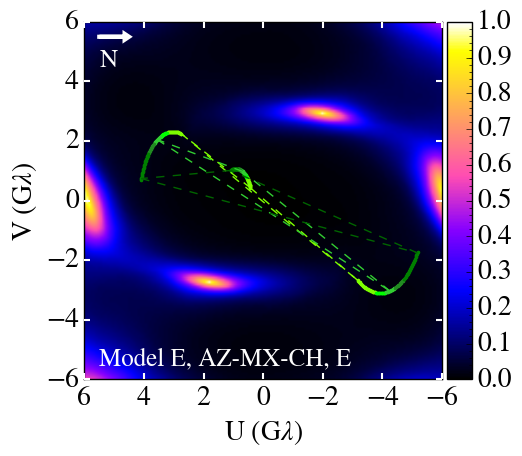}
\includegraphics[height=2in]{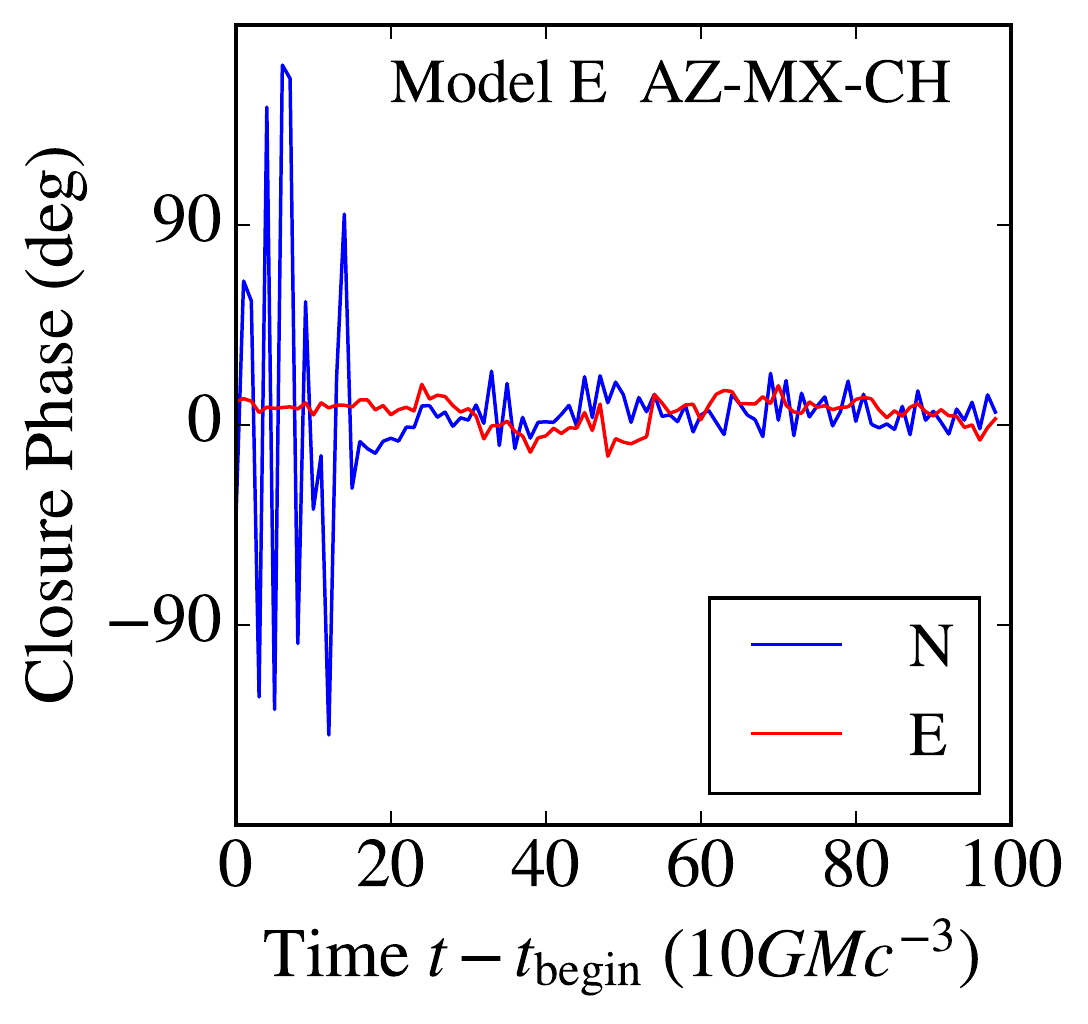}
\caption{The effect of Earth rotation on the variability of closure phases. The top two color maps show the directional dispersion of Model A, the green tracks and dashed triangles correspond to the HI-AZ-CA closure triangle for a black hole pointing South (left panel) and East (middle panel). 
The bottom two color maps show the directional dispersion of Model E, the green tracks and dashed triangles correspond to the AZ-MX-CH closure triangle for a black hole pointing North (left panel) and East (middle panel). 
During the course of an observation both closure triangles move from light green to dark green.
The rightmost column shows how closure phase varies as a function of time due to the combined effect of intrinsic variability from the simulation and the motion of the closure triangles shown in the color maps. 
Depending on the orientation of the black hole, the rotation of the Earth may move the triangles to regions of high variability of closure phases during an observation.
}
\label{fig:move_tri1}
\end{figure*}

We show an example of this situation in Figure~\ref{fig:move_tri1} for the small
HI-AZ-CA triangle (top panels) and the SANE model A as well as for the larger
AZ-MX-CH triangle (bottom panels) and for MAD model E, for two different orientations
of the black-hole spin. In two of the configurations shown (Model A, HI-AZ-CA South 
orientation and Model E AZ-MX-CH East orientation), the
closure phase remains very stable throughout the observation, because
the triangles remain away from the locations of the amplitude
minima. In a third configuration (Model A, East orientation), the
HI-AZ-CA triangle can follow Sgr~A* for $\simeq 4$~hr. Because the
size of this baseline track is comparable to the extent of the
high-variability region, the closure phase is variable throughout the
observation. In the last configuration (model E, North orientation),
only a part of the longer ($\simeq 6$~hr) baseline track cuts through
the region of high variability, causing a very sudden decline in the
closure phase variability in the midst of the observation.

\section{Discussion and Conclusions} \label{sec:conclusion}

\begin{figure}[t!]
\centering
\includegraphics[width=3.5in]{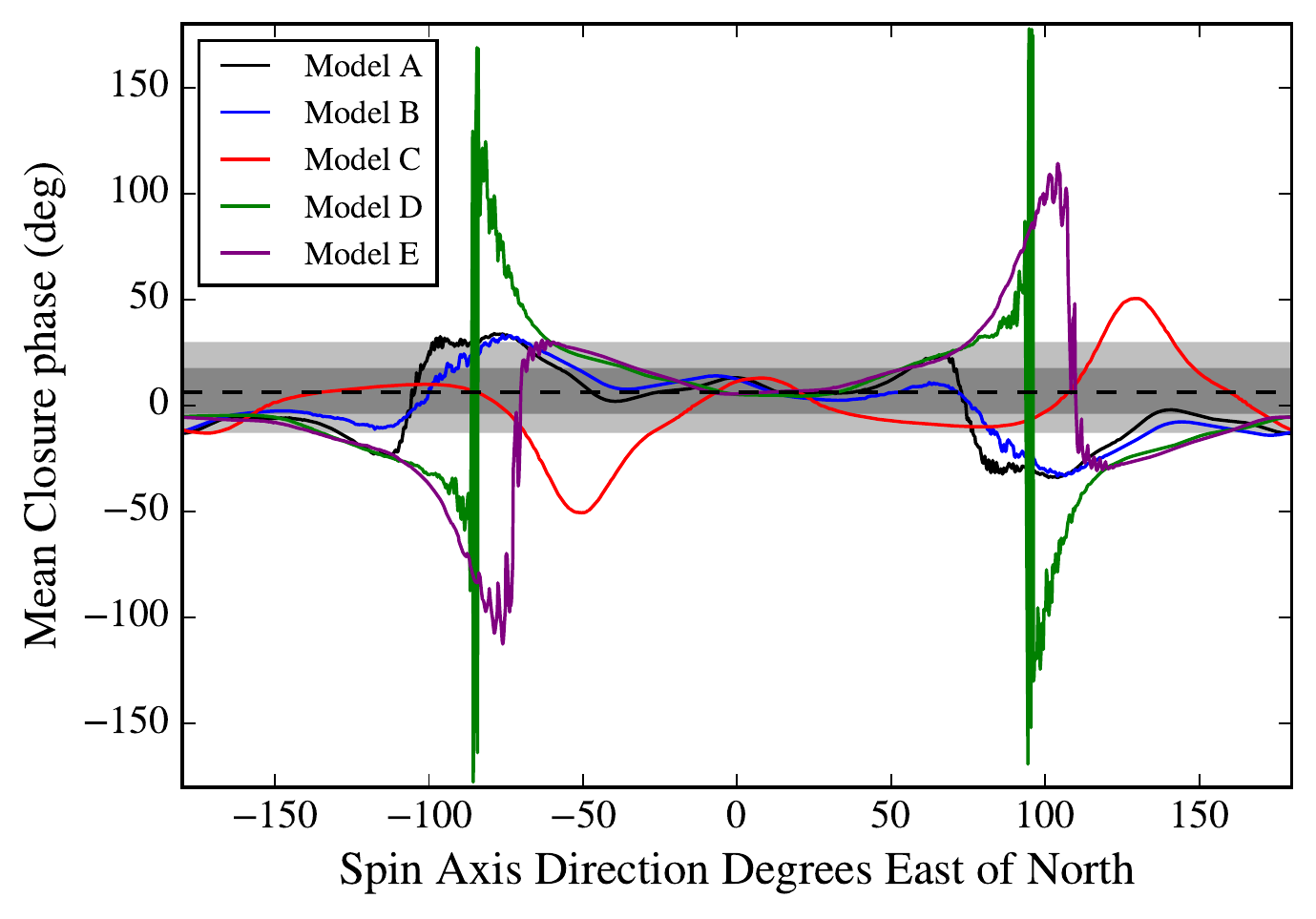}
\caption{Directional mean closure phase for the HI-AZ-CA triangle (in degrees) as a function of the orientation of the black hole spin axis. The dashed black line corresponds to the median closure phase measured by the EHT for this triangle \citep{2016ApJ...820...90F}.
The colored bands correspond to the ranges within which $68\%$ and $95\%$ of the measurements fall (neglecting statistical errors).
\\
\\
\\
\\
\\
\\
\\
}
\label{fig:orient_compare}
\end{figure}

We used five GRMHD+radiative transfer simulations of accretion onto
Sgr~A* to explore the predicted magnitudes of closure phases and their
variability for the upcoming interferometric observations with the
Event Horizon Telescope. We now compare these predictions to existing data to asses the prospects of distinguishing between different models and black hole spin orientations. 
Currently, there exist only limited
measurements of the closure phases, spanning different epochs, along
the HI-AZ-CA triangle. These yield a median value of $6.3_{-2.0}^{+0.7 \circ}$ \citep{2016ApJ...820...90F}. 
However, quantifying the exact magnitude of this variability requires a better understanding of calibration uncertainties than what is currently available. For this reason, we plot in Figure~\ref{fig:orient_compare} not only the median measured closure phase but also two horizontal bands that correspond to the ranges within which $68\%$ and $95\%$ of the 181 closure phase measurements fall.
Even though both the disk-dominated
SANE models and the jet-dominated MAD models we analyzed here have
significant asymmetric structures, Figure~\ref{fig:orient_compare} shows that they produce
closure phases and dispersions (Figure \ref{fig:orient}) in the HI-AZ-CA triangle that are consistent with the
measurements for a wide range of black-hole spin orientations on the
sky. 

In the near future, closure phases will be detected with the full EHT
array over a wide range of baseline triangles, covering long tracks in
the $u-v$ plane. Our models show that, for triangles with size similar
to that of the existing measurements, the closure phases will show little
variability, unless one of the baseline vertices crosses a region of
low visibility amplitude. However, the turbulent nature of the flow
introduces significant variability on the small scales and, hence,
significant closure phase variability might be present at large baseline triangles.

Despite this overall trend, the jet-dominated MAD models
that we studied produce less closure phase variability on the large
triangles than the disk-dominated SANE models. This is because the
images of the former are dominated by emission at the footpoints of
the jets and, even though these footpoints flicker, their image
structure is not greatly influenced by the variability in the
turbulent accretion flow. 
Therefore, future data will help distinguish between these possibilities. 
Furthermore, for both SANE and MAD models, we find that there is no trend between flaring events (see, e.g., Figure 1 in \citealt{2016arXiv160106799M}) and higher closure phase variability.

Our results have important implications for the image reconstruction
techniques that will rely on the closure phase data. Because of
the possibility of substantial dispersion, even at small triangles,
large amounts of high-quality data will need to be used to
characterize the variability properties of the closure phases. Image
reconstruction techniques will then need to take explicitly into
account the observed variability. Alternatively, if image
reconstruction techniques are used that rely on the assumption of a
stationary image, the regions of high closure phase variability will
need to be excised.

\acknowledgements
L.M. acknowledges support from NSF GRFP grant DGE~1144085.
C.K.C., D.P., and F.O. were partially supported by NASA/NSF TCAN award
NNX14AB48G and NSF grants Chandra Award No. TM6-17006X and AST~1312034.
D.M. acknowledges support from NSF grant AST-1207752.
All ray tracing calculations were performed with the \texttt{El~Gato}
GPU cluster at the University of Arizona that is funded by NSF award
1228509.


\appendix
\section{\\Appendix A: Directional Statistics}
Calculating statistical moments of distributions of quantities that are periodic in nature, such as closure phases, requires special care.
\citet{book} explore meaningful ways of determining the mean and dispersion of distributions of angles. Specifically, they suggest that the mean of a distribution of $n$ angles $\theta_j$ may be obtained using
\begin{eqnarray}
\bar{\theta}=
\begin{cases}
\tan^{-1}\left( \bar{S}/\bar{C}\right),& \text{if } \bar{C}\geq 0\\
\tan^{-1}\left( \bar{S}/\bar{C}\right)+\pi,              &\text{if } \bar{C}< 0,
\end{cases}
\end{eqnarray}
where
\begin{eqnarray}
\bar{S} = &&\frac{1}{n}\sum^n_{j=1} \sin (\theta_j)\\
\bar{C} = &&\frac{1}{n}\sum^n_{j=1} \cos (\theta_j).
\end{eqnarray}
In words, to calculate the mean of a distribution of angles, we calculate the mean of the unit vectors that correspond to the distribution of angles. 
The dispersion of a distribution of angles then is defined as
\begin{eqnarray}
D = \frac{1}{n}\sum^n_{j=1} \{  1-\cos(\theta_j-\bar{\theta})\}.
\end{eqnarray}
Hereafter, we will refer to this dispersion relation as the directional dispersion when comparing it to the dispersion relation commonly used for non-directional data.

To understand the behavior of the directional dispersion, we consider here two limiting cases.
For small deviation from the mean $(\theta_j-\bar{\theta})$, the directional dispersion can be approximated as
\begin{eqnarray}
D &&\simeq \frac{1}{n}\sum^n_{j=1} \left[  1-\left(1- \frac{(\theta_j-\bar{\theta})^2}{2}\right)\right]\nonumber\\
&&=\frac{1}{2n}\sum^n_{j=1} \left [ (\theta_j-\bar{\theta})^2\right]\nonumber\\
&&=\frac{\sigma^2}{2},
\end{eqnarray}
where we have denoted the normal definition of dispersion by $\sigma$.

In the limit of a continuous flat distribution of deviations from the mean, on the other hand, we find 
\begin{eqnarray}
D &&\simeq \frac{1}{2\pi}\int^{\pi}_{-\pi}[  1-\cos(\theta_j-\bar{\theta})]d\theta\nonumber\\
&&= 1-\frac{1}{2\pi}\int^{\pi}_{-\pi}\cos(\theta_j-\bar{\theta})d\theta\nonumber\\
&&= 1.
\end{eqnarray}

\begin{figure}[t!]
\centering
\includegraphics[height=3in]{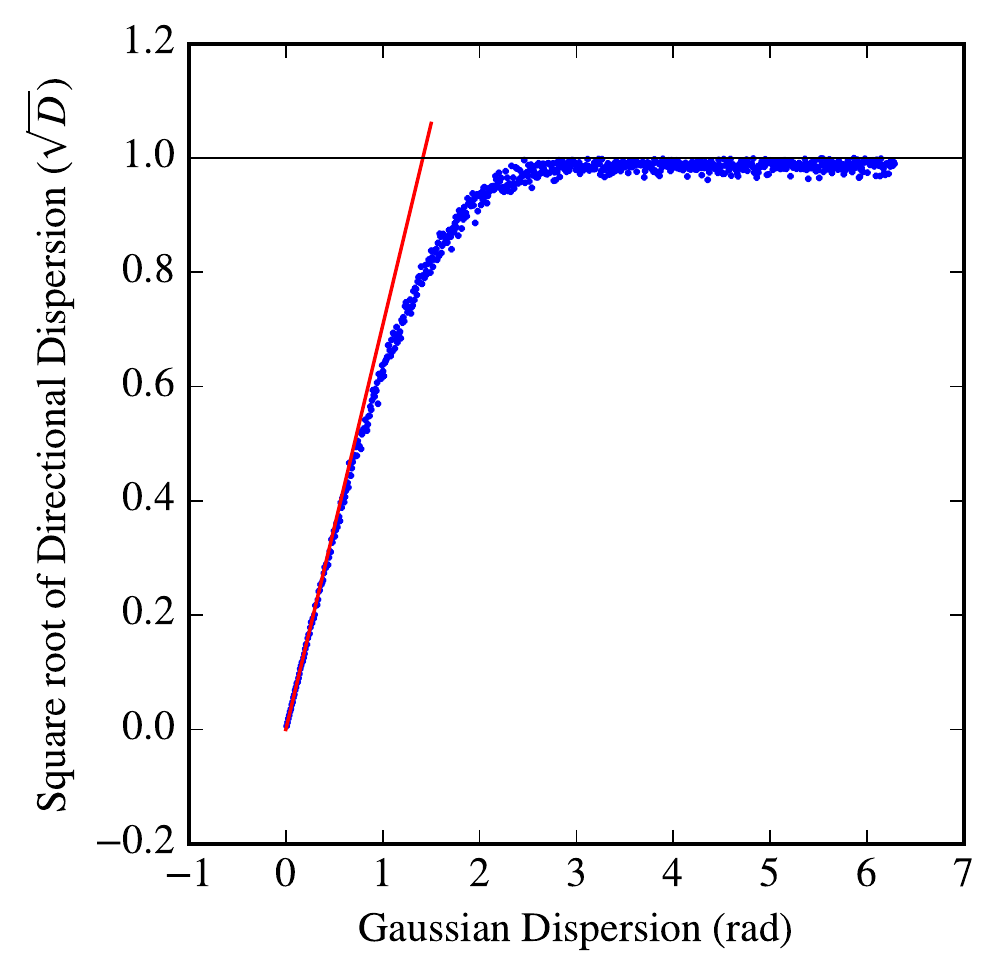}
\caption{Directional dispersion, $D$, of a Monte Carlo distribution of data calculated using directional statistics as a function of the standard deviation of a Gaussian distribution that was used to create the Monte Carlo data. 
The red line corresponds to $D=\sigma^2/2$. 
The horizontal black line is at $D=1$.
As the standard deviation in the Monte Carlo simulation reaches $\approx\pi$, the directional dispersion approaches unity and remains constant as the standard deviation in the Monte Carlo increases further.}
\label{fig:circ_disp}
\end{figure}

In Figure \ref{fig:circ_disp}, we explore the behavior of the directional dispersion further by comparing this quantity to the dispersion $\sigma$ of an ensemble of Monte Carlo points drawn from a Gaussian distribution. 
The blue points are the square root of the directional dispersion of simulated data created using Gaussian distributions with different dispersions.
As expected, the directional dispersion scales with the dispersion of the Gaussian distribution when the latter is small. 
However, as $\sigma \gtrsim \pi$, the periodic nature of the angular data causes the directional dispersion to asymptote to unity.

\bibliography{my,closure_phase}

\begin{thebibliography}{}
\expandafter\ifx\csname natexlab\endcsname\relax\def\natexlab#1{#1}\fi

\bibitem[{{Baganoff} {et~al.}(2001){Baganoff}, {Bautz}, {Brandt}, {Chartas},
  {Feigelson}, {Garmire}, {Maeda}, {Morris}, {Ricker}, {Townsley}, \&
  {Walter}}]{2001Natur.413...45B}
{Baganoff}, F.~K., {Bautz}, M.~W., {Brandt}, W.~N., {et~al.} 2001, \nat, 413,
  45

\bibitem[{{Broderick} {et~al.}(2011){Broderick}, {Fish}, {Doeleman}, \&
  {Loeb}}]{2011ApJ...738...38B}
{Broderick}, A.~E., {Fish}, V.~L., {Doeleman}, S.~S., \& {Loeb}, A. 2011, \apj,
  738, 38

\bibitem[{{Broderick} {et~al.}(2016){Broderick}, {Fish}, {Johnson},
  {Rosenfeld}, {Wang}, {Doeleman}, {Akiyama}, {Johannsen}, \&
  {Roy}}]{2016ApJ...820..137B}
{Broderick}, A.~E., {Fish}, V.~L., {Johnson}, M.~D., {et~al.} 2016, \apj, 820,
  137

\bibitem[{{Chan} {et~al.}(2013){Chan}, {Psaltis}, \&
  {{\"O}zel}}]{2013ApJ...777...13C}
{Chan}, C.-k., {Psaltis}, D., \& {{\"O}zel}, F. 2013, \apj, 777, 13

\bibitem[{{Chan} {et~al.}(2015{\natexlab{a}}){Chan}, {Psaltis}, {{\"O}zel},
  {Medeiros}, {Marrone}, {S{\c a}dowski}, \& {Narayan}}]{chan2015}
{Chan}, C.-k., {Psaltis}, D., {{\"O}zel}, F., {et~al.} 2015{\natexlab{a}},
  \apj, 812, 103

\bibitem[{{Chan} {et~al.}(2015{\natexlab{b}}){Chan}, {Psaltis}, {{\"O}zel},
  {Narayan}, \& {Sa{\c d}owski}}]{2015ApJ...799....1C}
{Chan}, C.-K., {Psaltis}, D., {{\"O}zel}, F., {Narayan}, R., \& {Sa{\c
  d}owski}, A. 2015{\natexlab{b}}, \apj, 799, 1

\bibitem[{{Dexter} {et~al.}(2010){Dexter}, {Agol}, {Fragile}, \&
  {McKinney}}]{2010ApJ...717.1092D}
{Dexter}, J., {Agol}, E., {Fragile}, P.~C., \& {McKinney}, J.~C. 2010, \apj,
  717, 1092

\bibitem[{{Do} {et~al.}(2009){Do}, {Ghez}, {Morris}, {Yelda}, {Meyer}, {Lu},
  {Hornstein}, \& {Matthews}}]{2009ApJ...691.1021D}
{Do}, T., {Ghez}, A.~M., {Morris}, M.~R., {et~al.} 2009, \apj, 691, 1021

\bibitem[{{Doeleman} {et~al.}(2009{\natexlab{a}}){Doeleman}, {Agol}, {Backer},
  {Baganoff}, {Bower}, {Broderick}, {Fabian}, {Fish}, {Gammie}, {Ho}, {Honman},
  {Krichbaum}, {Loeb}, {Marrone}, {Reid}, {Rogers}, {Shapiro}, {Strittmatter},
  {Tilanus}, {Weintroub}, {Whitney}, {Wright}, \&
  {Ziurys}}]{2009astro2010S..68D}
{Doeleman}, S., {Agol}, E., {Backer}, D., {et~al.} 2009{\natexlab{a}}, in ArXiv
  Astrophysics e-prints, Vol. 2010, astro2010: The Astronomy and Astrophysics
  Decadal Survey

\bibitem[{{Doeleman} {et~al.}(2009{\natexlab{b}}){Doeleman}, {Fish},
  {Broderick}, {Loeb}, \& {Rogers}}]{2009ApJ...695...59D}
{Doeleman}, S.~S., {Fish}, V.~L., {Broderick}, A.~E., {Loeb}, A., \& {Rogers},
  A.~E.~E. 2009{\natexlab{b}}, \apj, 695, 59

\bibitem[{{Doeleman} {et~al.}(2002){Doeleman}, {Phillips}, {Rogers},
  {Attridge}, {Titus}, {Smythe}, {Cappallo}, {Buretta}, {Whitney}, {Krichbaum},
  {Graham}, {Alef}, {Polatidis}, {Bach}, {Witzel}, {Zensus}, {Greve},
  {Grewing}, {Freund}, {Strittmatter}, {Ziurys}, {Wilson}, {Fagg}, \&
  {Gay}}]{2002evn..conf..223D}
{Doeleman}, S.~S., {Phillips}, R.~B., {Rogers}, A.~E.~E., {et~al.} 2002, in
  Proceedings of the 6th EVN Symposium, ed. E.~{Ros}, R.~W. {Porcas}, A.~P.
  {Lobanov}, \& J.~A. {Zensus}, 223

\bibitem[{{Doeleman} {et~al.}(2008){Doeleman}, {Weintroub}, {Rogers},
  {Plambeck}, {Freund}, {Tilanus}, {Friberg}, {Ziurys}, {Moran}, {Corey},
  {Young}, {Smythe}, {Titus}, {Marrone}, {Cappallo}, {Bock}, {Bower},
  {Chamberlin}, {Davis}, {Krichbaum}, {Lamb}, {Maness}, {Niell}, {Roy},
  {Strittmatter}, {Werthimer}, {Whitney}, \& {Woody}}]{2008Natur.455...78D}
{Doeleman}, S.~S., {Weintroub}, J., {Rogers}, A.~E.~E., {et~al.} 2008, \nat,
  455, 78

\bibitem[{{Dolence} {et~al.}(2009){Dolence}, {Gammie}, {Mo{\'s}cibrodzka}, \&
  {Leung}}]{2009ApJS..184..387D}
{Dolence}, J.~C., {Gammie}, C.~F., {Mo{\'s}cibrodzka}, M., \& {Leung}, P.~K.
  2009, \apjs, 184, 387

\bibitem[{{Fish} {et~al.}(2011){Fish}, {Doeleman}, {Beaudoin}, {Blundell},
  {Bolin}, {Bower}, {Chamberlin}, {Freund}, {Friberg}, {Gurwell}, {Honma},
  {Inoue}, {Krichbaum}, {Lamb}, {Marrone}, {Moran}, {Oyama}, {Plambeck},
  {Primiani}, {Rogers}, {Smythe}, {SooHoo}, {Strittmatter}, {Tilanus}, {Titus},
  {Weintroub}, {Wright}, {Woody}, {Young}, \& {Ziurys}}]{2011ApJ...727L..36F}
{Fish}, V.~L., {Doeleman}, S.~S., {Beaudoin}, C., {et~al.} 2011, \apjl, 727,
  L36

\bibitem[{{Fish} {et~al.}(2016){Fish}, {Johnson}, {Doeleman}, {Broderick},
  {Psaltis}, {Lu}, {Akiyama}, {Alef}, {Algaba}, {Asada}, {Beaudoin},
  {Bertarini}, {Blackburn}, {Blundell}, {Bower}, {Brinkerink}, {Cappallo},
  {Chael}, {Chamberlin}, {Chan}, {Crew}, {Dexter}, {Dexter}, {Dzib}, {Falcke},
  {Freund}, {Friberg}, {Greer}, {Gurwell}, {Ho}, {Honma}, {Inoue}, {Johannsen},
  {Kim}, {Krichbaum}, {Lamb}, {Le{\'o}n-Tavares}, {Loeb}, {Loinard},
  {MacMahon}, {Marrone}, {Moran}, {Mo{\'s}cibrodzka}, {Ortiz-Le{\'o}n},
  {Oyama}, {{\"O}zel}, {Plambeck}, {Pradel}, {Primiani}, {Rogers}, {Rosenfeld},
  {Rottmann}, {Roy}, {Ruszczyk}, {Smythe}, {SooHoo}, {Spilker}, {Stone},
  {Strittmatter}, {Tilanus}, {Titus}, {Vertatschitsch}, {Wagner}, {Wardle},
  {Weintroub}, {Woody}, {Wright}, {Yamaguchi}, {Young}, {Young}, {Zensus}, \&
  {Ziurys}}]{2016ApJ...820...90F}
{Fish}, V.~L., {Johnson}, M.~D., {Doeleman}, S.~S., {et~al.} 2016, \apj, 820,
  90

\bibitem[{{Fraga-Encinas} {et~al.}(2016){Fraga-Encinas}, {Mo{\'s}cibrodzka},
  {Brinkerink}, \& {Falcke}}]{2016A&A...588A..57F}
{Fraga-Encinas}, R., {Mo{\'s}cibrodzka}, M., {Brinkerink}, C., \& {Falcke}, H.
  2016, \aap, 588, A57

\bibitem[{{Gammie} {et~al.}(2003){Gammie}, {McKinney}, \&
  {T{\'o}th}}]{2003ApJ...589..444G}
{Gammie}, C.~F., {McKinney}, J.~C., \& {T{\'o}th}, G. 2003, \apj, 589, 444

\bibitem[{{Genzel} {et~al.}(2003){Genzel}, {Sch{\"o}del}, {Ott}, {Eckart},
  {Alexander}, {Lacombe}, {Rouan}, \& {Aschenbach}}]{2003Natur.425..934G}
{Genzel}, R., {Sch{\"o}del}, R., {Ott}, T., {et~al.} 2003, \nat, 425, 934

\bibitem[{{Ghez} {et~al.}(2008){Ghez}, {Salim}, {Weinberg}, {Lu}, {Do}, {Dunn},
  {Matthews}, {Morris}, {Yelda}, {Becklin}, {Kremenek}, {Milosavljevic}, \&
  {Naiman}}]{2008ApJ...689.1044G}
{Ghez}, A.~M., {Salim}, S., {Weinberg}, N.~N., {et~al.} 2008, \apj, 689, 1044

\bibitem[{{Gillessen} {et~al.}(2009){Gillessen}, {Eisenhauer}, {Trippe},
  {Alexander}, {Genzel}, {Martins}, \& {Ott}}]{2009ApJ...692.1075G}
{Gillessen}, S., {Eisenhauer}, F., {Trippe}, S., {et~al.} 2009, \apj, 692, 1075

\bibitem[{{Jennison}(1958)}]{1958MNRAS.118..276J}
{Jennison}, R.~C. 1958, \mnras, 118, 276

\bibitem[{{Johannsen} {et~al.}(2012){Johannsen}, {Psaltis}, {Gillessen},
  {Marrone}, {{\"O}zel}, {Doeleman}, \& {Fish}}]{2012ApJ...758...30J}
{Johannsen}, T., {Psaltis}, D., {Gillessen}, S., {et~al.} 2012, \apj, 758, 30

\bibitem[{{Johnson} \& {Gwinn}(2015)}]{2015ApJ...805..180J}
{Johnson}, M.~D., \& {Gwinn}, C.~R. 2015, \apj, 805, 180

\bibitem[{{Lu} {et~al.}(2016){Lu}, {Roelofs}, {Fish}, {Shiokawa}, {Doeleman},
  {Gammie}, {Falcke}, {Krichbaum}, \& {Zensus}}]{2016ApJ...817..173L}
{Lu}, R.-S., {Roelofs}, F., {Fish}, V.~L., {et~al.} 2016, \apj, 817, 173

\bibitem[{Mardia \& Jupp(1999)}]{book}
Mardia, K., \& Jupp, P. 1999, Directional Statistics, 1st edn. (Baffins Lane,
  Chichester, West Sussex, PO19 IUD England: Wiley)

\bibitem[{{Marrone} {et~al.}(2008){Marrone}, {Baganoff}, {Morris}, {Moran},
  {Ghez}, {Hornstein}, {Dowell}, {Mu{\~n}oz}, {Bautz}, {Ricker}, {Brandt},
  {Garmire}, {Lu}, {Matthews}, {Zhao}, {Rao}, \& {Bower}}]{2008ApJ...682..373M}
{Marrone}, D.~P., {Baganoff}, F.~K., {Morris}, M.~R., {et~al.} 2008, \apj, 682,
  373

\bibitem[{{Medeiros} {et~al.}(2016){Medeiros}, {Chan}, {Ozel}, {Psaltis},
  {Kim}, {Marrone}, \& {Sadowski}}]{2016arXiv160106799M}
{Medeiros}, L., {Chan}, C.-k., {Ozel}, F., {et~al.} 2016, ArXiv e-prints,
  arXiv:1601.06799

\bibitem[{{Mo{\'s}cibrodzka} {et~al.}(2011){Mo{\'s}cibrodzka}, {Gammie},
  {Dolence}, \& {Shiokawa}}]{2011ApJ...735....9M}
{Mo{\'s}cibrodzka}, M., {Gammie}, C.~F., {Dolence}, J.~C., \& {Shiokawa}, H.
  2011, \apj, 735, 9

\bibitem[{{Narayan} {et~al.}(2012){Narayan}, {S{\c a}dowski}, {Penna}, \&
  {Kulkarni}}]{sadowski_fix}
{Narayan}, R., {S{\c a}dowski}, A., {Penna}, R.~F., \& {Kulkarni}, A.~K. 2012,
  \mnras, 426, 3241

\bibitem[{{Neilsen} {et~al.}(2013){Neilsen}, {Nowak}, {Gammie}, {Dexter},
  {Markoff}, {Haggard}, {Nayakshin}, {Wang}, {Grosso}, {Porquet}, {Tomsick},
  {Degenaar}, {Fragile}, {Houck}, {Wijnands}, {Miller}, \&
  {Baganoff}}]{2013ApJ...774...42N}
{Neilsen}, J., {Nowak}, M.~A., {Gammie}, C., {et~al.} 2013, \apj, 774, 42

\bibitem[{{Porquet} {et~al.}(2008){Porquet}, {Grosso}, {Predehl}, {Hasinger},
  {Yusef-Zadeh}, {Aschenbach}, {Trap}, {Melia}, {Warwick}, {Goldwurm},
  {B{\'e}langer}, {Tanaka}, {Genzel}, {Dodds-Eden}, {Sakano}, \&
  {Ferrando}}]{2008A&A...488..549P}
{Porquet}, D., {Grosso}, N., {Predehl}, P., {et~al.} 2008, \aap, 488, 549

\bibitem[{{S{\c a}dowski} {et~al.}(2013){S{\c a}dowski}, {Narayan}, {Penna}, \&
  {Zhu}}]{2013MNRAS.436.3856S}
{S{\c a}dowski}, A., {Narayan}, R., {Penna}, R., \& {Zhu}, Y. 2013, \mnras,
  436, 3856

\bibitem[{{Shiokawa} {et~al.}(2012){Shiokawa}, {Dolence}, {Gammie}, \&
  {Noble}}]{2012ApJ...744..187S}
{Shiokawa}, H., {Dolence}, J.~C., {Gammie}, C.~F., \& {Noble}, S.~C. 2012,
  \apj, 744, 187

\end{thebibliography}
\end{document}